\documentclass[fleqn,usenatbib]{mnras}

\usepackage{newtxtext}%,newtxmath}
\usepackage[T1]{fontenc}
\usepackage{ae,aecompl}

\usepackage{graphicx}	% Including figure files
\usepackage{amsmath}	% Advanced maths commands
\usepackage{amssymb}	% Extra maths symbols

\defcitealias{parsons16}{paper~I}
\defcitealias{Rebassa-Mansergas17}{paper~II}
\defcitealias{lagosetal20-1}{paper~III}
\defcitealias{Ren20}{paper~V}
\defcitealias{Hernandez21}{paper~IV}
\title[Two PCEBs with {\it TESS} light curve]{The White Dwarf Binary Pathways Survey VI: two close post common envelope binaries with {\it TESS} light curves}

\author[M.-S. Hernandez et al.]{
M.S. Hernandez$^{1}$,\thanks{E-mail: mercedes.hernandez@postgrado.uv.cl}
M.R. Schreiber$^{2,3}$,
S.G. Parsons$^{4}$,
B.T. G\"{a}nsicke$^{5,6}$,
O. Toloza$^{2,5}$,\newauthor
G. Tovmassian$^{7}$,
M. Zorotovic$^{1}$,
F. Lagos$^{1,3,8}$,
R. Raddi$^{9}$,
A. Rebassa-Mansergas$^{9,10}$,\newauthor
J.J. Ren$^{11}$,
C. Tappert$^{1}$
\\
$^{1}$Instituto de F{\'i}sica y Astronom{\'i}a de la Universidad de Valpara{\'i}so, Av. Gran Breta\~na 1111, Valpara{\'i}so, Chile.\\
$^{2}$Departamento de F{\'i}sica, Universidad T\'ecnica Federico Santa Mar\'ia, Av. España 1680, Valpara{\'i}so, Chile.\\
$^{3}$Millennium Nucleus for Planet Formation, NPF, Valpara{\'i}so, Chile\\
$^{4}$Department of Physics and Astronomy, University of Sheffield, Sheffield S3 7RH, UK.\\
$^{5}$University of Warwick, Department of Physics, Gibbet Hill Road, Coventry, CV4 7AL, United Kingdom.\\
$^{6}$Centre for Exoplanets and Habitability, University of Warwick, Coventry CV4 7AL, UK.\\
$^{7}$Instituto de Astronom{\'i}a, Universidad Nacional Aut{\'o}noma de M{\'e}xico, Apartado Postal 877, Ensenada, Baja California, 22800, M{\'e}xico.\\
$^{8}$European Southern Observatory (ESO), Alonso de Cordova 3107, Vitacura, Santiago, Chile\\
$^{9}$Departament de F{\'i}sica, Universitat Polit{\`e}cnica de Catalunya, c/Esteve Terrades 5, E-08860 Castelldefels, Spain.\\
$^{10}$Institute for Space Studies of Catalonia, c/Gran Capit{\`a} 2-4, Edif. Nexus 201, 08034 Barcelona, Spain.\\
$^{11}$Key Laboratory of Space Astronomy and Technology, National Astronomical Observatories, Chinese Academy of Sciences,\\ Beijing 100101, P. R. China.\\
}

\date{Accepted 2022 March 02. Received 2022 March 02; in original form 2021 August 22}

\pubyear{2019}

\begin{document}
\label{firstpage}
\pagerange{\pageref{firstpage}--\pageref{lastpage}}
\maketitle

\begin{abstract}
%The white dwarf binary pathways survey is dedicated to 
Establishing a large sample of post common envelope binaries (PCEBs) that consist of a white dwarf plus an intermediate mass companion star of spectral type AFGK, offers the potential to provide new constraints on theoretical models of white dwarf binary formation and evolution.
Here we present a detailed analysis of two new systems, TYC\,110-755-1 and TYC\,3858-1215-1. % which brings the number of well characterized PCEBs with secondary stars of spectral type AFGK to 13. 
Based on radial velocity measurements we find the orbital periods of the two systems to be $\sim 0.85$ and $\sim 1.64$\,days, respectively. In addition,  {\it HST}  spectroscopy of TYC\,110-755-1 allowed us to measure the mass of the white dwarf in this system ($0.78 \,\mathrm{M}_{\odot}$). 
We furthermore analysed {\it TESS} high time resolution photometry and find both secondary stars to be magnetically extremely active. Differences in the photometric and spectroscopic periods of TYC\,110-755-1 indicate that the secondary in this system is  differentially rotating. 
Finally, studying the past and future evolution of both systems, we conclude that the common envelope efficiency is likely similar in close white dwarf plus AFGK binaries and PCEBs with M-dwarf companions and find a wide range of possible evolutionary histories for both systems. While TYC\,3858-1215-1 will run into dynamically unstable mass transfer that will cause the two stars to merge and evolve into a single white dwarf,  TYC\,110-755-1 is a progenitor of a cataclysmic variable system with an evolved donor star. 
%Our results further confirm the diversity of evolutionary pathways of close white dwarf binaries, demonstrate that {\it TESS} light curves can be used to identify short period systems among our target sample, and that the secondary stars in WD+AFGK close binaries
\end{abstract}

%%%%%%%%%%%%%%%%%%%%%%%%%%%%%%%%%%%%%%%%%%%%%%%%%%

%%%%%%%%%%%%%%%%% BODY OF PAPER %%%%%%%%%%%%%%%%%%

\section{Introduction}

Close white dwarf (WD) binaries are the result of the evolution of main sequence star binaries. After the more massive star evolves into a giant star it may fill its Roche lobe and the resulting mass transfer is typically dynamically unstable. As a consequence the entire binary evolves through a 
common envelope (CE) \citep[e.g.,][]{Taam10, Passy11, Ivanova13}. During this phase, angular momentum and orbital energy are extracted from the orbit and unbind the envelope \citep{webbink84-1, zorotovicetal10-1}, revealing a close binary containing a white dwarf and its main-sequence companion. These systems, known as post-common-envelope binaries (PCEBs), are thought to be the progenitors of some of the most exotic objects in the Galaxy, such as cataclysmic variables \citep{warner1995CV}, AM CVn binaries \citep{warner1995AM}, hot subdwarf stars \citep{Heber2009}, supersoft X-ray sources \citep[SSSs,][]{KahabkaHeuvel1997}, and double degenerates binaries \citep{webbink84-1}. 
The latter two are progenitor candidates for type Ia supernovae \citep[SNe Ia,][]{bildsten2007,WangZhanwen2012}. The zoo of possible evolutionary outcomes demonstrate the complexity of close white dwarf binary evolution.  
%\vspace{0.2cm}
%The population of  %``\textit{Post-common envelope binaries from SDSS}''
%has been studies in great detail in the last decades.   
%

Important progress towards our understanding of close compact binary star evolution has been derived from surveys of PCEBs \citep{alberto07,Rebassa-Mansergas12,Nebot11}. 
%which allowed to 
%The orbital period distribution of 
%close white dwarf binaries with M dwarf companions 
%systems was measured \citep{Nebot11}
%which allowed to 
These surveys provided constraints on the common envelope efficiency
\citep{zorotovicetal10-1} and the subsequent angular momentum loss due to magnetic braking \citep{schreiber10-1}. 
However, these works were entirely focused on PCEBs composed of a white dwarf and a low-mass (M\,<\,0.4\,M$_\odot$) main-sequence (spectral type M) companion. If angular momentum loss drives the systems close enough to initiate the mass transfer from the main-sequence star to the white dwarf, this mass transfer will be stable given the small mass ratio, i.e. virtually all PCEBs with M dwarf companion will become CVs sooner or later which then may evolve towards shorter orbital periods or merge due to consequential angular momentum loss \citep{schreibertal16-1}.
In other words, previous surveys of PCEBs do not provide constraints on the formation mechanism of objects as important as
SSSs and double degenerates which must form from PCEBs with secondary stars more massive than M dwarfs. 

The main aim of {\it the white dwarf binary pathways survey project} is to extend the search for PCEBs towards systems containing secondary stars of A, F, G or K spectral type (WD+AFGK).  A sufficiently large sample of this type of systems will provide new constraints on the close white dwarf binary formation. Additionally, they  provide potentially large implications for our understanding of close white dwarf binary formation and evolution, including the two main channels that are thought to produce a type Ia supernovae.

We established a large target sample ($\sim 1600$) of WD+AFGK binary stars 
combining ultraviolet ({\rm UV}) observations from  {\em GALEX} with optical wavelengths observations from the Radial Velocity Experiment \citep[RAVE,][]{kordopatisetal13-1} in \citet[][hereafter paper I]{parsons16}, the Large Sky Area Multi-Object Fibre Spectroscopic Telescope \citep[LAMOST,][]{dengetal12-1} in \citet[][hereafter paper II]{Rebassa-Mansergas17} and more recently with the  {\em Tycho-Gaia} Astrometric Solution \citep[TGAS,][]{Lindegren16} described in \citet[hereafter paper V]{Ren20}.
%Several observational campaigns have been performed  over the last five years to obtain high-resolution spectroscopy of our selected targets. First, with the aim of identifying those systems that show radial velocity variations, and then to follow them up in order to measure their orbital periods. 
So far, this search has resulted in the discovery of TYC\,6760-497-1, the first known pre-SSS binary \citep{parsons15}, followed by TYC\,7218-934-1, a triple system whose discovery led us to estimate the fraction of triple systems that could contaminate the WD+AFGK sample to be less than 15 per cent \citep[][hereafter paper III]{lagosetal20-1}. Finally, we recently showed that rather similar close white dwarf binaries with G-type star companions can have quite distinctive futures \citep[][hereafter paper IV]{Hernandez21}. 

Among the systems we identified is one object that will likely evolve into a SSS.  As  SSSs are difficult to observe  due to their short lifetime and because the X-ray emission is hard to detect as it is easily absorbed by neutral hydrogen in the galactic plane, observing progenitors of  SSSs might represent our best chance to provide observational constraints on this evolutionary channel. We also found a system that will evolve into a CV with an evolved donor. These systems are thought to be one of the possible progenitors of AM~CVn stars with an estimated birthrate through this channel of ($0.5-1.3$)$ \times 10^{-3} yr^{-1}$ in our Galaxy \citep{Podsiadlowski03}. The third system we found will lead to a dynamical merger of the WD and the secondary star when the latter is on the asymptotic giant branch (AGB). The combination of the extreme mass ratio and the convective envelope of the giant star causes unstable mass transfer onto the WD driving to the formation of a common envelope. The merger of the WD with the massive degenerate core of the AGB star may lead to a SN\,Ia explosion through the so--called core degenerate scenario which could produce $\sim1-2$ per cent of these events \citep{Zhou15}.
 
Here we present two additional close binaries discovered by our project, TYC\,110-755-1 and TYC\,3858-1215-1 (hereafter TYC\,110 and TYC\,3858, respectively), composed of a white dwarf with a G and K type secondary star, respectively, and with orbital periods shorter than two days. % We found that TYC\,3858 have a third not interactive component.    We obtained Hubble Space Telescope ({\it HST})  spectroscopy for TYC\,110, which further confirmed that our selection algorithms is very efficient (so far all 11 systems observed with  {\it HST}  turned out to indeed contain a white dwarf) and which allowed to measure the mass of the white dwarf.   

\section{Observations}
White dwarfs with AFGK spectral type are much more difficult to identify in observational surveys than WD+M binaries as the main sequence companion star dominates the emission at optical wavelength. The target selection of {\em{The White Dwarf Binary Pathways Survey}} is therefore based on combining information from optical surveys with {\textit{GALEX}} data \citepalias[for details see][]{Rebassa-Mansergas17,Ren20}.
%{\bf{Here we need to describe the target selection which does not appear anywhere else in the paper!!!}}
In this paper we present two of the selected targets that turned out to be close white dwarf binaries: TYC\,110 and TYC\,3858. 
We characterize both systems combining photometric and spectroscopic observations and 
in what follows we describe in detail the performed observations and data reduction. 

\subsection{Optical high resolution spectroscopy}\label{sec:highspec}
%\section{Observations and data reduction}

After identifying TYC\,110 and TYC\,3858 as PCEB candidates from TGAS and LAMOST samples \citepalias[][respectively]{Ren20, Rebassa-Mansergas17}, we performed  time-resolved high-resolution optical spectroscopic follow up observations to confirm their close binarity by means of radial velocity variations. Thanks to our targets being bright ($V=10.5$\,mag  and $11.48$\,mag, for TYC\,110 and TYC\,3858 respectively)  the observations were performed under poor weather conditions.

Observations with three different high-resolution spectrographs installed at different telescopes were obtained for both systems. We used the Echelle SPectrograph from REosc for the Sierra San pedro martir Observatory \citep[ESPRESSO,][]{levin95} at the 2.12~m telescope of the Observatorio Astron\'omico Nacional at San Pedro M\'artir  (OAN-SPM)\footnote{http://www.astrossp.unam.mx}, M\'exico and observed with the Ultraviolet and Visual Echelle Spectrograph \citep[UVES,][]{Dekker00} on the Very Large Telescope of the European Southern Observatory (ESO) in Cerro Paranal. Additional data were obtained with the High Resolution Spectrograph \citep[HRS,][]{Jiang02} at the Xinglong 2.16~m telescope (XL216) located in China.

ESPRESSO provides spectra covering the wavelength range from $3500$ to $7105$~\AA \, with a spectral resolving power of R $\approx 18\,000$ for a slit width of  150~$\mu$m. A Th-Ar lamp before each exposure was used for wavelength calibration. 
HRS provides spectra of a $49\,800$ resolving power for a fixed slit width of 0.19~mm and covers a wavelength range of $\sim 3650$--$10\,000$~\AA. Th-Ar arc spectra were taken at the beginning and end of each night. The ESPRESSO and HRS spectra were reduced using the \textit{echelle} package in IRAF\footnote{IRAF is distributed by the National Optical Astronomy Observatories, which are operated by the Association of Universities for Research in Astronomy, Inc., under cooperative agreement with the National Science Foundation.}.
Standard procedures, including bias subtraction, cosmic-ray removal, and wavelength calibration, were carried out using the corresponding tasks in IRAF. 
The observations carried out with UVES have a spectral resolution of 58\,000 for a 0.7--arcsec slit. With its two-arms, UVES covers the wavelength range of $3000$--$5000$~\AA\, (blue) and $4200$--$11\,000$~\AA\, (red), centered at 3900 and 5640\,\AA\, respectively. Standard reduction was performed making use of the specialized pipeline EsoReflex workflow \citep{Freudling13}.  

\subsection{HST spectroscopy}

The presence of a white dwarf in our WD+AFGK binary star candidates is inferred from {\rm UV} excess emission in  {\em GALEX}. We have previously observed ten candidate systems with the \textit{Hubble Space Telescope} \citepalias{parsons16} and in all cases the presence of a white dwarf was confirmed which indicates that our selection algorithm is very efficient and that our target sample is rather clean.   
However, excess {\rm UV} emission or optical spectroscopy does not provide solid constraints on the mass of the white dwarf. 
%discovered by 
%At optical wavelength 
%In the case of the WDs in binary systems with AFGK type companions, the secondary star outshines the WD at optical wavelengths. Therefore, 
The only way to reliably measure the white dwarf mass is through ultraviolet spectroscopy.% and fortunately we observed one of the two binaries presented in this work, i.e. TYC\,110, 
%with {\it HST} .  

We obtained far-ultraviolet ({\rm FUV}) spectroscopy of TYC\,110 using the Space Telescope Imaging Spectrograph \citep[STIS,][]{Kimble98} on-board the {\it HST} (GO 16224) on 2021 January 31. Spectra were obtained over one spacecraft orbit, resulting in a spectrum with total exposure time of 2185\,seconds. They were acquired with the MAMA detector and the G140L grating. The spectra cover a wavelength range of 1150--1730\,\AA\, with a resolving power between 960--1440.  
The {\rm FUV} spectra were reduced and wavelength calibrated following the standard STIS pipeline \citep{Sohn19}.  The {\it HST} spectrum contains emission lines of He\,II at 1640\,\AA\, and C\,IV at 1550\,\AA, and shows the broad Ly-alpha absorption of a hydrogen white dwarf atmosphere.

%\subsection{Binary parameters}

%To progress with the understanding of close WD binary star formation and evolution, we first  established a large sample of post-common envelope binaries with early type (AFGK) secondary stars. The second and most critical step is to measure the orbital period of the binary in order to constrain the rest of the parameters of the system. It also plays an important role to determine the past and future of a given PCBE.

\subsection{ {\it TESS} photometry}\label{sec:tess}

Motivated by the idea to  measure the main sequence star rotational periods, and resolve if the secondary stars in close binary stars are indeed tidally locked,% which is particularly important for TYC\,3858 as the $v\sin{i}$ method is our only estimate of the white dwarf mass, 
we extracted high cadence light curves for both systems from the \textit{TESS} database \citep{Ricker15}. The \textit{TESS} mission splits observations into 26 overlapping 24x96 degree sky sectors over the Northern and Southern hemispheres; and each sector was observed for approximately one month. The light curves were downloaded from the Mikulski Archive for Space Telescopes (MAST\footnote{https://mast.stsci.edu}) web service. TYC\,110 data belong to sectors 5 and 32, while TYC\,3858 was observed in sectors 15, 16, 22 and 23.  The time covered by each sector can be found in Table\,\ref{tab:LCperiods}. Pre-search Data Conditioned Simple Aperture Photometry (PDCSAP) flux was used for this study.

\section{Stellar and orbital parameters}

The performed observations allow us to characterize the stellar and orbital parameters of both systems. We describe in this 
section the characterization of the secondary star and the determination of the orbital period from our high resolution spectroscopic observations and {\it TESS} photometry. The white dwarf parameters for TYC\,110 are then determined from our {\it HST} observations.

\subsection{The secondary stars} 
\label{sec:secondary}

\begin{figure}
	\includegraphics[width=\columnwidth]{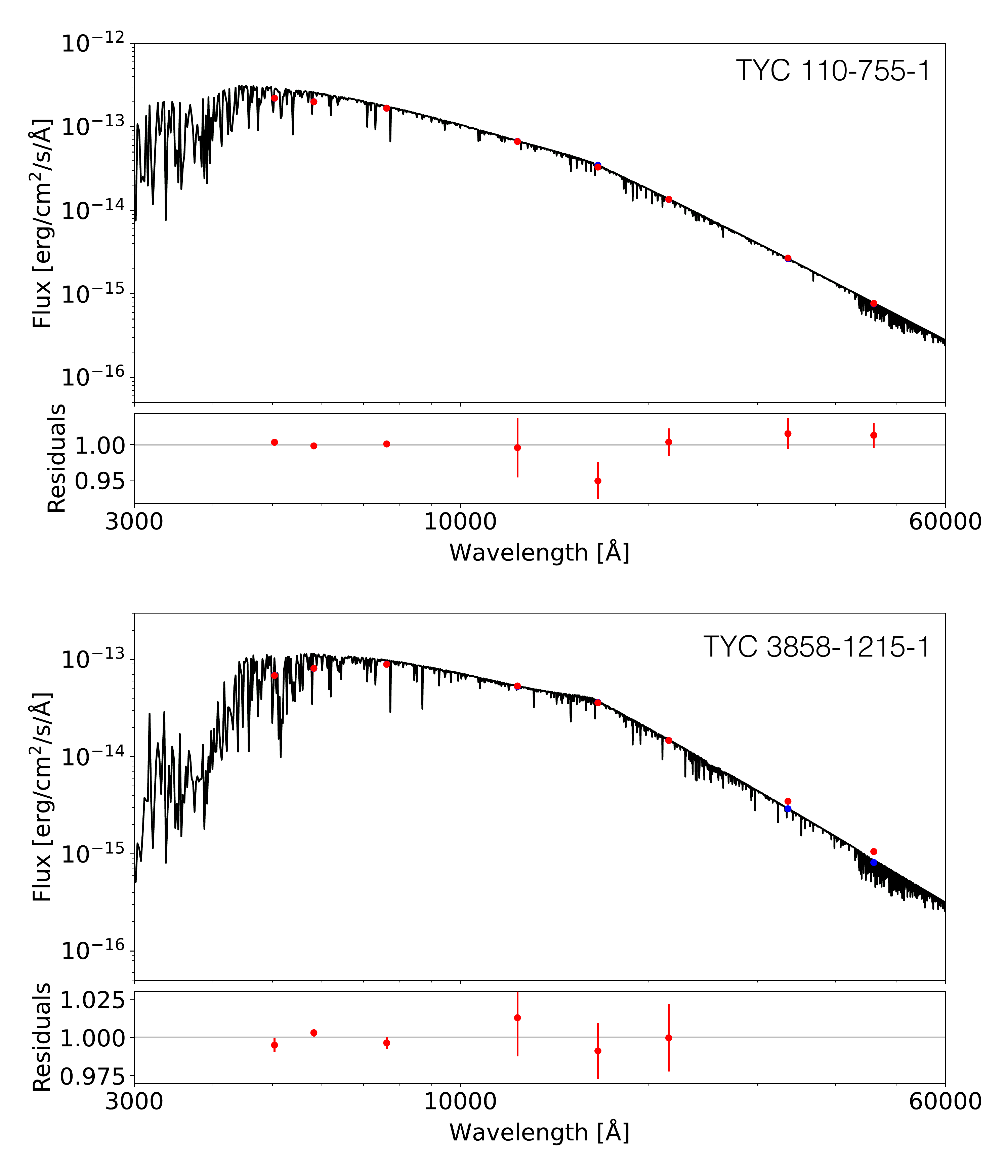}
    \caption{Spectral energy distributions created from $G_\mathrm{BP}$, $G_\mathrm{RP}$ and $G$ {\it Gaia} bands, $J$, $H$ and $K_s$ band data from {\rm{2MASS}}, and $W1$ and $W2$ band data from WISE (red dots) were fitted with MARCS.GES theoretical spectra (black line). }
    \label{fig:SED_ms}
\end{figure}

\begin{figure}
	\includegraphics[width=\columnwidth]{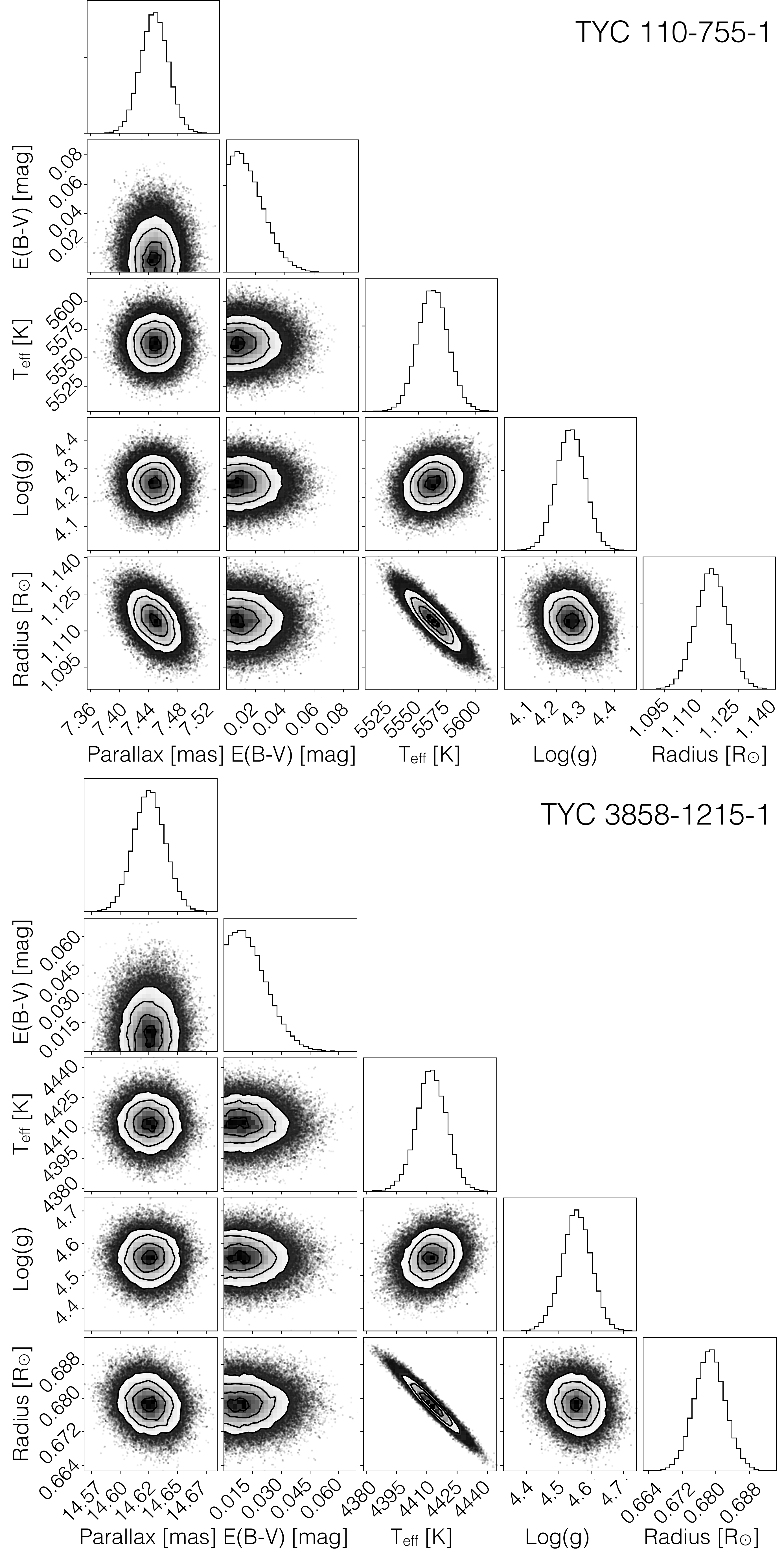}
    \caption{Posterior probability distributions for model parameters obtained from the SED fit.}
    \label{fig:corner}
\end{figure}

To determine the stellar parameters of the main-sequence star in our binaries, normalized spectra with the highest signal-to-noise ratio  of each target were fitted to MARCS.GES\footnote{https://marcs.astro.uu.se/} models \citep{Gustafsson08} using iSpec \citep{Blanco2014}, allowing to measure the effective temperatures ($T\mathrm{_{eff}}$), surface gravities ($\log{g}$), metallicities ($Z$) and rotational broadening ($v\sin{i}$) for our targets. The spectral fitting procedure started with different initial values. On the one hand we used the Gaia DR2 \citep{Gaia18} $T\mathrm{_{eff}}$ values, $\log{g}$=4.5\,dex, $v\sin{i}$=10\,km\,s$^{-1}$ and solar metallicity. On the other hand, fits were also performed with initial values varied by $\pm$100\,K in $T\mathrm{_{eff}}$, using $\log{g}$=4.0\,dex and 5.0, metallicities of $Z=-1$ and +0.5\,dex, and $v\sin{i}$=1 and 100\,km\,s$^{-1}$ in order to ensure that all fits converge to the same solution. 

The resulting best-fit model was combined
with the spectral energy distribution (SED) of the target scaled to  {\it Gaia} EDR3 \citep{Gaia20} parallaxes  (7.44(01)\,mas for TYC\,110 and 14.62(01)\,mas for TYC\,3858) to estimate the radius of the secondary star. 
The SED of our targets were created using the {\it Gaia} EDR3 $G_\mathrm{BP}$, $G_\mathrm{RP}$ and $G$ bands, along with $J$, $H$ and $K_s$ band data  from {\rm{2MASS}} \citep{Cutri03}, and $W1$ and $W2$ band data from {\em WISE} \citep{Cutri12}.  It is worth highlighting that the photometry for TYC\,3858 is affected by several nearby stars, rendering the $W1$ and $W2$ band data unreliable for this object because of the spatial resolution, so they were not included in the fit. The {\it Gaia} and {\rm{2MASS}} data appear to be unaffected by this.   
The SED fit uses the parallax, reddening,  $T\mathrm{_{eff}}$, $\log{g}$ and radius as initial parameters for the Markov chain Monte Carlo (MCMC) method \citep{Press07} to determine the final values and their uncertainties.  $T\mathrm{_{eff}}$ and $\log{g}$ are initialised based on the values from the spectral fit, while in the case of the parallax and reddening we use the priors and parameter ranges constrained by {\it Gaia} EDR3 \citep{Lallement19} and the STILISM reddening map \footnote{https://stilism.obspm.fr/} \citep{Capitanio17} which provide uncertainties. In any case, given the distance and location of our targets we expect reddening to be negligible. The radius is initialised at a sensible value for a main sequence star with the corresponding $\log{g}$ and $T\mathrm{_{eff}}$. 
The production chain used 50 walkers, each with 10\,000 points. The first 1500 points were classified as a "burn-in", and were removed from the final results. 
The best fit values and their uncertainties are listed in Table \ref{tab:parameters}. 
 For both stars, the obtained stellar parameters do not correlate with the small reddening,  and the obtained reddening is even consistent with zero, which makes sense given the high latitude and the close distance to these targets and which is in agreement with the value provided by the {\it Gaia/2MASS} 3D extinction map 
\footnote{https://astro.acri-st.fr/gaia\_dev/}. 
The obtained SEDs and parameter distributions of both targets are plotted in Fig.\,\ref{fig:SED_ms} and \ref{fig:corner}, respectively.

\subsection{Radial velocities}\label{sec:maths} % used for referring to this section from elsewhere

Radial velocities (RVs) from ESPRESSO and UVES spectra were computed using the cross-correlation technique against a binary mask representative of a G-type star. The uncertainties in radial velocity were computed using scaling relations \citep[][]{jordan14} with the signal-to-noise ratio and the width of the cross-correlation peak, which was calibrated with Monte Carlo simulations.

Radial velocities from the spectra collected with the HRS spectrograph were obtained by fitting the normalised Ca\,II absorption triplet (at $8498.02, 8542.09$, and $8662.14$~\AA) with a combination of a second order polynomial and a triple-Gaussian profile of fixed separations, as described in \citet{rebassa-mansergasetal17-1}.
The radial velocity uncertainty is obtained by summing the fitted error and a systematic error of 0.5~km~s$^{-1}$ in quadrature \citepalias{Ren20}. All radial velocities and observing details obtained for TYC\,110 and TYC\,3858 are provided in table \ref{tab:RV_tyc110} and \ref{tab:RV_tyc3858} of the appendix.

 \begin{figure}
	\includegraphics[width=\columnwidth]{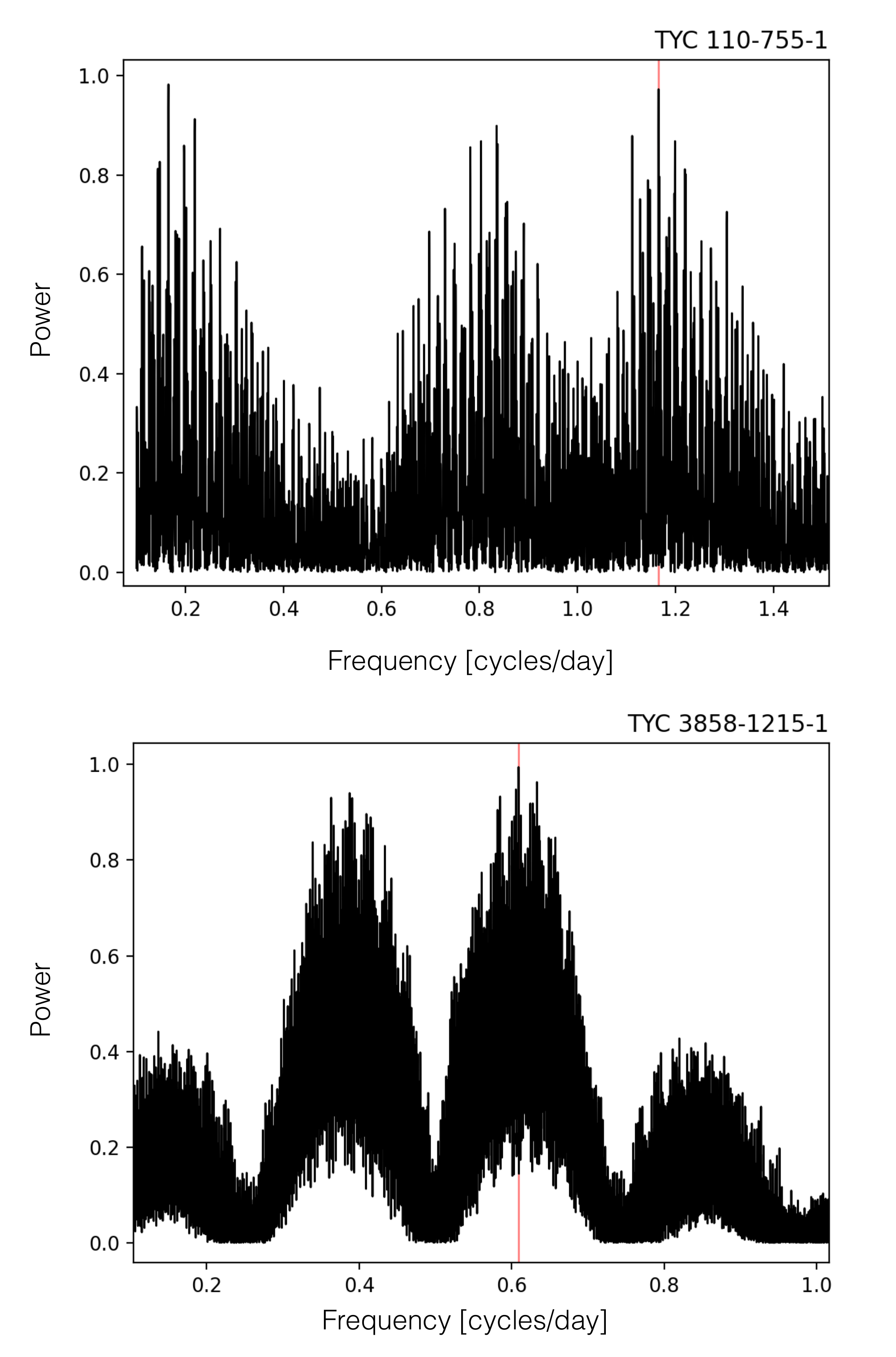}
    \caption{ Lomb-Scargle periodograms  based on the spectroscopic observations of TYC\,110 (top) and TYC\,3858 (bottom). The red lines highlight the highest peak which corresponds to the orbital period of the systems.}
    \label{fig:periodogramRVs}
\end{figure}

%In  Generalized Lomb-Scargle Periodogram (GLS, \citet{Zechmeister09}). 
We used the least-squares spectral analysis method based on the Generalized Lomb-Scargle Periodogram \citep[GLS][]{Zechmeister09}, %\citep{Lomb76,Scargle82} 
implemented in the astroML python library \citep{VanderPlas12}, to identify the orbital period of both systems. Additionally, a specialized Monte Carlo sampler created to find converged posterior samplings for Keplerian orbital parameters called {\em{The Joker}} \citep{Price-Whelan17, Price-Whelan20} was used. The Joker is especially useful for non-uniform data and also allows to identify eccentric orbits. As we have previously identified a system with an eccentric orbit which turned out to be a triple star system with a white dwarf component \citepalias{lagosetal20-1}, double-checking circular orbits derived from Lomb-Scargle periodograms was absolutely required, especially for TYC\,3858 which is a hierarchical triple system where the PCEB is the inner binary. The third object is an M-dwarf star at $\sim330$~au from the inner binary, far away enough to not have any significant impact on the ultraviolet detection or to affect on some way the evolution of the PCEB. We discovered the third object by searching the {\it Gaia} EDR3 for sources around the system with consistent parallaxes and proper motions.

\begin{figure}
	\includegraphics[width=\columnwidth]{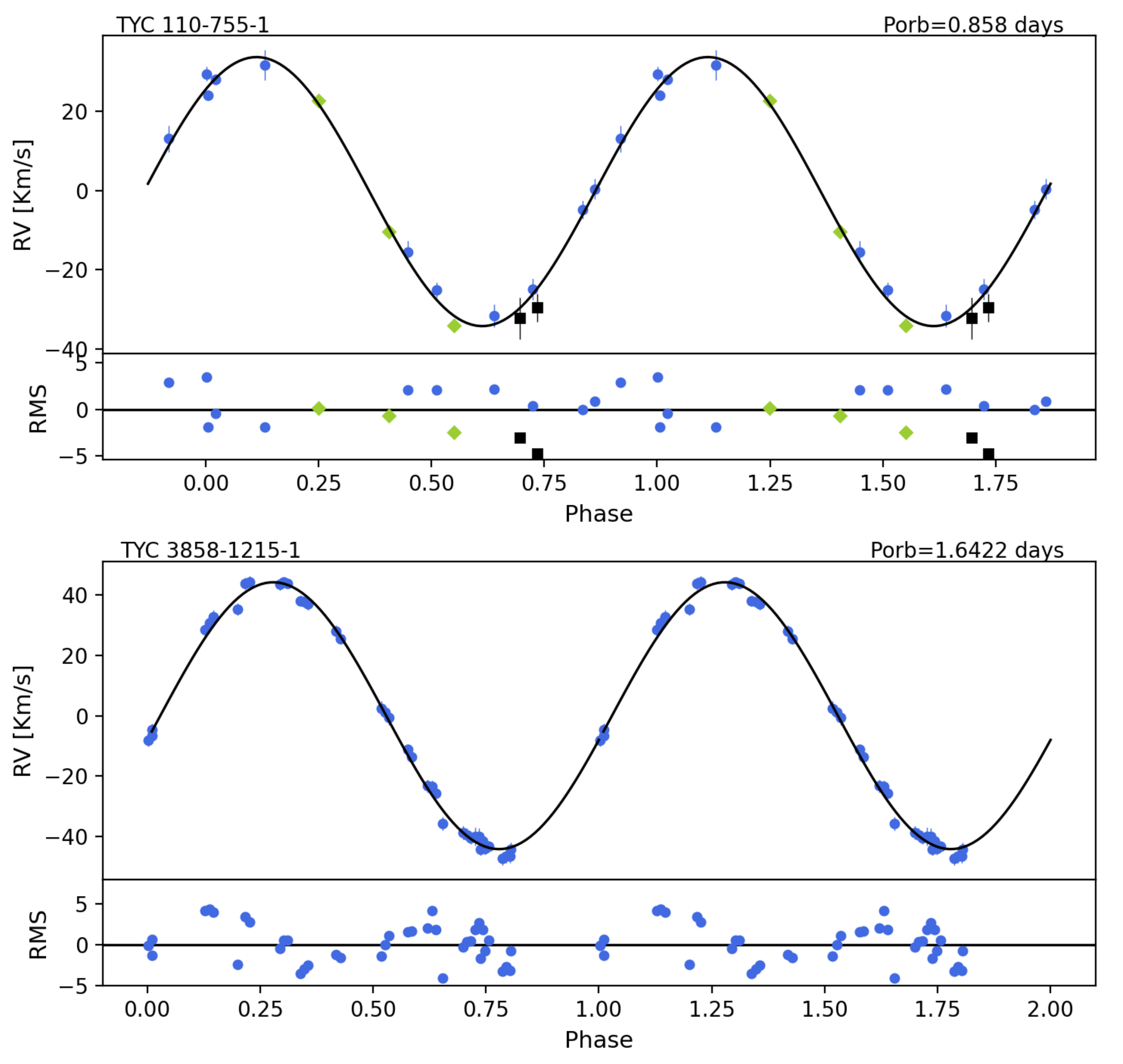}
    \caption{Phase-folded radial velocity curves for the main-sequence stars in TYC\,110 (top) and TYC\,3858 (bottom), with the corresponding residuals to the best fit. The colors indicate which instrument was used to obtain the measurement: OAN-SPM (blue circles), XL216 (black squares), and UVES (green diamonds).}
    \label{fig:RVC}
\end{figure}

The GLS periodograms  of the RV measurements for both systems are shown in Fig. \ref{fig:periodogramRVs}. %
Folding the radial velocity data with the best period yields the radial velocity curves shown in Fig.~\ref{fig:RVC}. Data from different telescopes are plotted in different colors. The measured orbital periods for TYC\,110 and TYC\,3858 are 0.858 and 1.6422~days, respectively. The eccentricity in both cases is negligible. 
%Both orbital period determination methods (the least-squares spectral analysis and {\em{The Joker}}) found 

While the best fit period provides a reasonable representation of the data, 
the periodograms show several aliases with similar powers to the selected period. 
We tested with the three highest peaks for each target and find that 
the selected period reproduces the data best but that the second highest peaks provide only a factor of 8 and 1.5 larger $\chi^2$ values for TYC\,110 and TYC\,3858, respectively. Our radial velocity data alone does therefore not provide an unambiguous orbital period measurement. However, as we shall see in the next section, {\it TESS} photometry available for both targets clearly breaks the degeneracy and confirms that the best fit periods are the correct orbital periods in both cases. 

%and found the
%spectroscopic periods 
%determined by our spectroscopic observations. However, as we shall see in the following section, {\it TESS} photometry allows to select the correct orbital period. %$\chi^2$ values resulting from sine fits are at least a factor $8$ and $1.5$ worse for  These relatively small difference do not 
%{\bf Additionally, we found for TYC\,110 the highest peak corresponding to a period of 6.056\,days that fits well with the data too. Therefore, we do not have enough information to decide a definite period only with radial velocities.}
%The aliases with similar power in the periodograms disagreed drastically with some radial velocity measurements, and the $\chi^2$ values resulting from sine fits using the periods corresponding to the second highest peaks in the periodograms are at least a factor $8$ and $1.5$ {\bf higher } , respectively, and can therefore be excluded.

%\section{{\it HST} ultraviolet spectroscopy}\label{sec:HST}

\subsection{Photometric observations}

 \begin{figure}
	\includegraphics[width=\columnwidth]{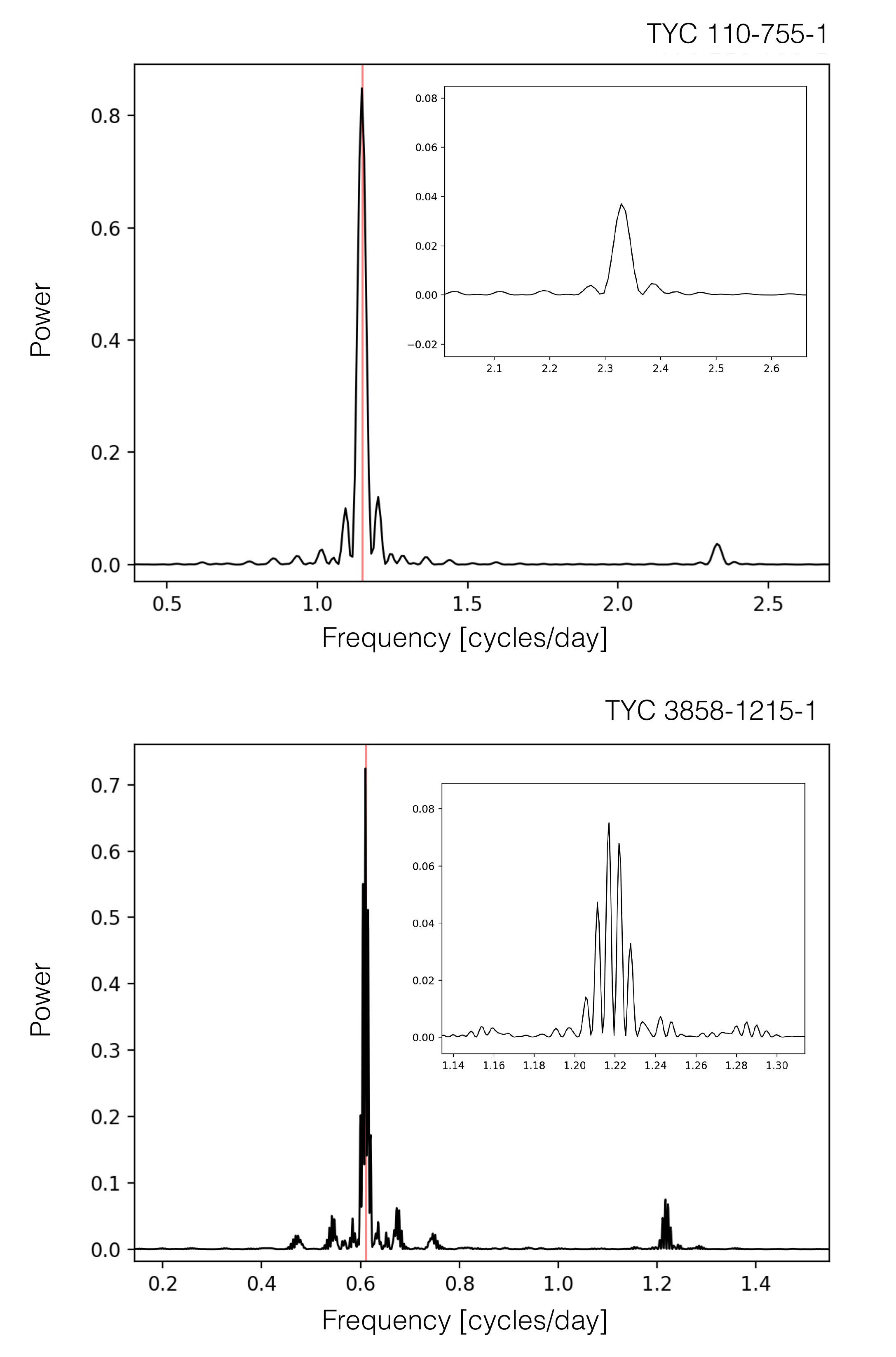}
    \caption{ The power spectra of TYC\,110 (top) and TYC\,3858 (bottom) calculated for their {\it TESS} light curves combining data from all sectors. The inserted zoom corresponds to the small range of frequencies around the second peak found at half the orbital period.}
    \label{fig:periodogramTESS}
\end{figure}

\begin{figure}
	\includegraphics[width=\columnwidth]{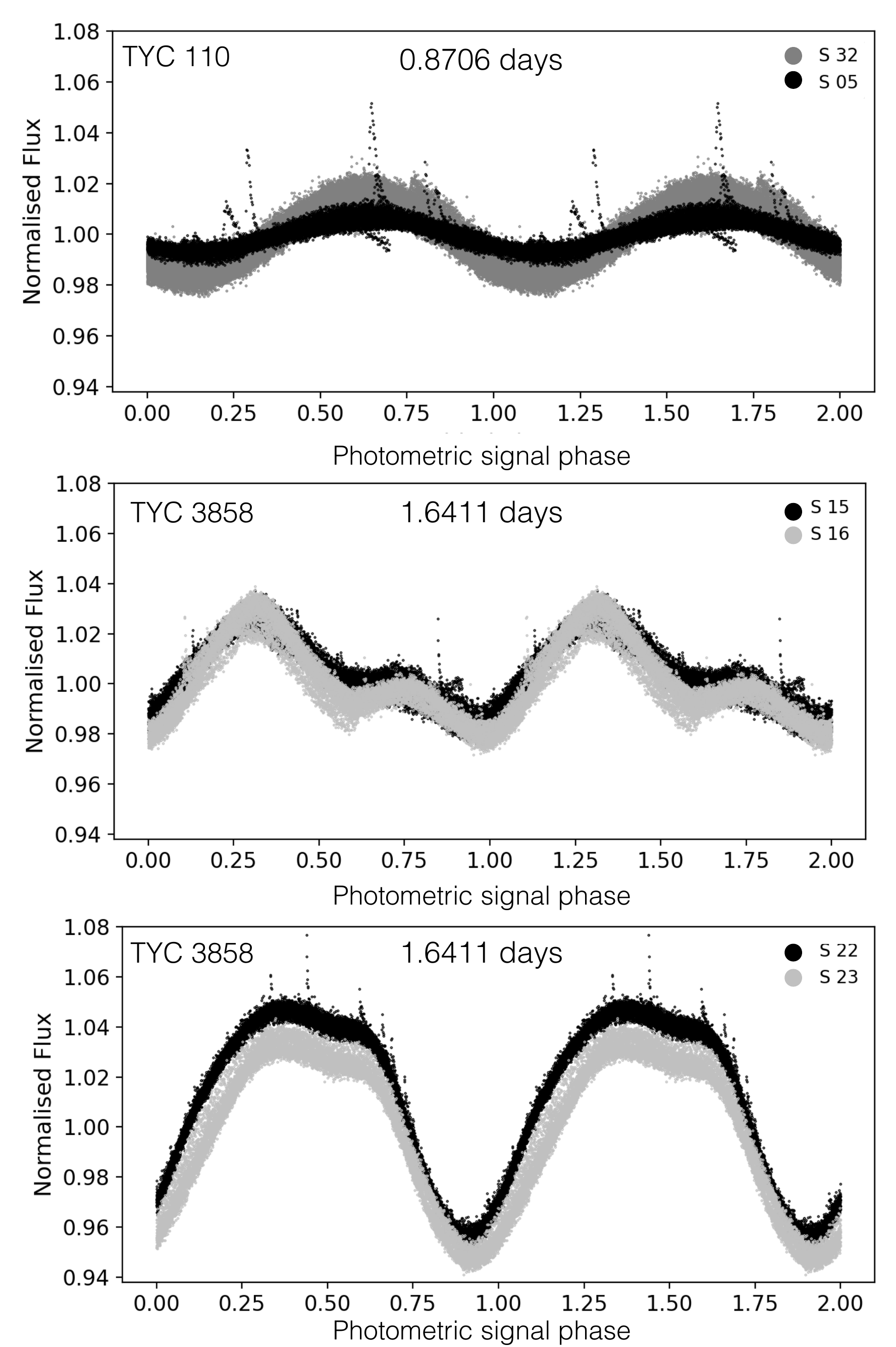}
    \caption{The upper panel shows the {\it TESS} light curve of TYC\,110 in sector 5 (black points) and  32 (grey points). The rotational period is consistent between the two sectors, but there is a change in the amplitude of the variations. We split the light curve of TYC\,3858 into two different plots to better visualise the changes in shape and amplitude over time. In the middle panel, observations show sector 15 (black) and 16 (grey) with an amplitude of  $\sim$1.7 percent. In the bottom panel, which shows data from sector 22 (black) and 23 (grey), the amplitude increases to $\sim$4.2 percent.
}
    \label{fig:LC}
\end{figure}

We applied the least-squares spectral analysis method based on the classical Lomb-Scargle periodogram \citep{Lomb76,Scargle82} to the \textit{TESS} data of both systems (see Fig.\,\ref{fig:periodogramTESS}). We also removed the dominant photometric period from both data sets and investigated the residuals. 
The phase folded light curve for both targets are shown in Fig.\,\ref{fig:LC}. We found the light curve of TYC\,110 is rather sinusoidal with short term variations that we interpret as flares. TYC\,3858 shows a double-peaked feature and also some flare activity. Shape and amplitude of the light curve change from one epoch to the other for both targets while the periodicity of the signal for the individual sectors are identical within the uncertainties. Table\,\ref{tab:LCperiods} shows the periods and amplitudes measured from each sector for both systems. In what follows we discuss the results obtained from the {\it TESS} light curves in the light of the spectroscopic period for both binary stars. 

%and TYC\,3858 secondary  and 1.6392 days, respectively. After removing the main period signal in each sector, TYC\,110 and TYC\,3858  and 0.8211\,days, respectively. These periods are exactly half the spectroscopic orbital period entirely consistent with ellipsoidal modulation of the main sequence stars. 
%only in the case of TYC\,110 which fills its Roche lobe by 66 percent. 
%TYC\,3858 fills its Roche lobe a maximum of 28 percent and the period we got is just an alias. Periodograms are shown in Fig.\,\ref{fig:periodogramTESS}. 
%We have to spot that TYC\,110, shows a difference of 18\,minutes between orbital and rotational period. These difference may be the main responsible that the WD mass obtained with the {\it  {\it HST} } spectra is slightly higher than the range of masses estimated with the $v\sin{i}$ method. {\bf Given that, for the latest, we assume the binary is tidally locked}. 

\begin{table}
    \centering
    \caption{List of the periods and amplitudes obtained by analyzing individual sectors of TESS observations of each system. The last column shows the flux that was used to normalize the corresponding light curve.}
    \label{tab:LCperiods}
    \resizebox{\columnwidth}{!}{\begin{tabular}{lllll}
    \hline
    Target/ & Time range & Period & Amplitude  & Flux zero  \\
    sector & [BJD] &[days] &  [$e^-$/s]  &[$e^-$/s] \\
    \hline 
    TYC\,110-755-1 &&&& \\
    05 & 2458437.99--2458463.94 &0.870$\pm$0.005   & 120.4$\pm$0.4  & 18309       \\
    32 & 2459174.23--2459199.97 &0.871$\pm$0.004   &  279.8$\pm$0.2  & 18640 \\
    \hline
    TYC\,3858-1215-1 &&&&\\
    15 & 2458711.36--2458735.69   & 1.65$\pm$0.07 & 160.7$\pm$0.6  & 9637 \\
    16 & 2458738.64--2458762.64    & 1.60$\pm$0.07 & 172.0$\pm$0.9  & 9606 \\
    22 & 2458899.89--2458926.49   & 1.64$\pm$0.02 & 206.5$\pm$0.8  & 9713  \\
    23 & 2458930.21--2458954.87   & 1.64$\pm$0.03 & 398.9$\pm$0.8  & 9584\\
    \hline
    \end{tabular}}
\end{table}

%\subsection{Differential rotation} 

%which can introduce some jitters into radial velocity measurements, and   

\subsubsection{TYC~110-755-1}

The main period we found  for TYC\,110 is $0.87066\pm 0.00099$\,days which is close to the best fit spectroscopic period with a difference of 18 minutes. On one hand this suggest that the period we selected from the radial velocities corresponds to the orbital period. On the other hand, the difference between the radial velocity period and the period obtained from the \textit{TESS} light curve is larger than the combined period uncertainties. This result is very puzzling at first glance as the general hypothesis is that for binaries with orbital periods shorter than 10 days synchronised rotation can be assumed for main sequence stars \citep{Fleming19}. 

\begin{figure}
	\includegraphics[width=\columnwidth]{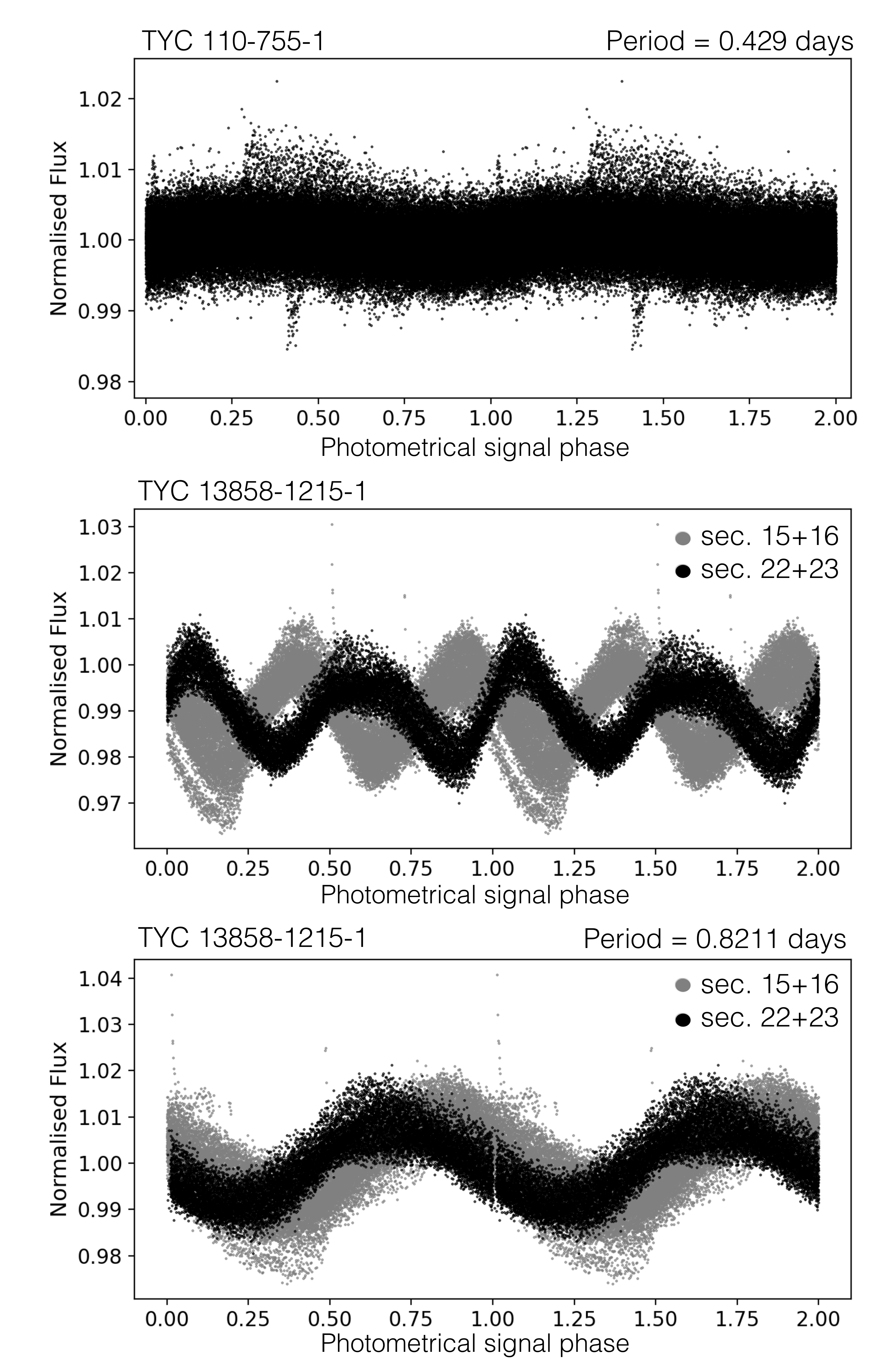}
    \caption{ Variation detected in the {\it TESS} light curves after removing the main photometric signal. Top: Ellipsoidal variation found in TYC\,110 (sector 32). The data is phase folded over a period 0.429\,days which corresponds to exactly half the orbital period obtained from our radial velocity measurements. Middle: Variation detected in TYC\,3858, sectors 15 $+$ 16 (gray) and 22 $+$ 23 (black) after removing the main signal and phase folded over the periods 0.82055 and 0.82202\,days, respectively.  Bottom:  Variation found in TYC\,3858 after a second round of subtracting a sinusoidal fit. The residuals are phase folded over a  period 0.8211\,days which corresponds to exactly half the orbital period obtained from our radial velocity measurements. The signals at half the orbital period are not caused by ellipsoidal variations in the case of TYC\,3858 because their amplitude largely exceeds the expected amplitude for ellipsoidal variations given the system parameters. Instead we suggest that spots at roughly opposite sides of the secondary star might cause the observed periodicities. }
    \label{fig:elipse}
\end{figure}

However, by further investigating the {\it TESS} light curve we found a possible solution for this problem. After removing the main periodic signal from  
the {\it TESS} light curve, we found a statistically significant signal with a period of $0.4290$\,days, i.e. exactly half the spectroscopic period (Fig.\,\ref{fig:periodogramTESS}). This signal is consistent with being caused by ellipsoidal variations given that according to the measured stellar masses and radius of the secondary, the latter is filling approximately 66 per cent of its Roche-lobe.  The resulting light curve is shown in Fig.\,\ref{fig:elipse}. The measured amplitude of the signal is $0.106\pm0.002$ percent of the measured flux. To further test our hypothesis that this signal corresponds to ellipsoidal variations, we calculate the theoretically predicted amplitude of ellipsoidal variations using the measured stellar and binary parameters of the system. 
According to \citet{Morris93} and \citet{Zucker07} the relative flux produced by ellipsoidal variations is given 
by 
\begin{equation}
    \frac{\delta F}{F}=0.15 \frac{(15+u_1)(1+\beta_1)}{(3-u_1)}\left(\frac{R_1}{a}\right)^3 q\sin^2{i},
    \label{eq:dff}
\end{equation}
where $\frac{\delta F}{F}$ is the fractional semi-amplitude of the ellipsoidal variation, star 1 refers to the main-sequence star, $u_1$ is the linear limb darkening coefficient and $\beta_1$ is the gravity darkening exponent for the main-sequence star in the TESS filter, which we obtained from tables 24 and 29\footnote{https://cdsarc.cds.unistra.fr/viz-bin/cat/J/A+A/600/A30\#/browse} of \citet{Claret17}. $R_1$ is the radius of the main-sequence star, $a$ is the semi-major axis, $q=M_{WD}/M_1$ is the mass ratio and $i$ is the system inclination.
%we make use of eq.\,\ref{eq:dff} to estimate what the amplitude of the ellipsoidal modulation should be.  
We found that using the parameter from Table\,\ref{tab:parameters} the theoretical predicted values for the amplitude of ellipsoidal variations for TYC\,110 cover the range of $0.137-0.172$ percent of the measured flux, which is in good agreement with the value measured from the light curve. This result confirms that the spectroscopic period is the orbital period of the binary star system and the signal at half the orbital period is most likely caused by ellipsoidal variations. 
 
To understand the difference in the rotational and orbital period of the secondary star, a closer look at the period variations over time is required. In Fig.\,\ref{fig:trailerphot} we show the residuals as a function of orbital cycle number and find 
that there clearly is a variation in the light curve that changes over time. 
We interpret this as being caused by differential rotation of the secondary star, i.e. while the equatorial regions are synchronised with the orbital motion, at higher latitudes this might not be the case, or vice versa. 
If a starspot is moving to a different latitude, which rotates at a slightly slower velocity, the spot comes into view out-of-time on each orbit, resulting in a longer period. The position of the starspots moving on the surface of the star may also cause the amplitude variation we observed in the light curve shown in Fig.\,\ref{fig:LC}.% as well as the general evolution of the starspot itself. 
This scenario is plausible as it is common for solar type stars to be differentially rotating. This has been seen in some donor stars in cataclysmic variables and other non-eclipsing binaries  \citep{Lurie17,Bruning06,Maceroni90}. %Our interpretation of the period difference as being caused by differential rotation 
%also nicely explains why the white dwarf 
%mass estimate based on $v\sin i$ is in disagreement with the mass measured from  {\it HST}  spectroscopy. A key assumption of the  
%$v\sin i$ method is synchronised rotation of the entire secondary star and therefore, if the star is in fact differentially rotating, the mass estimate becomes incorrect.  
Finally, TYC\,110 clearly shows flares in its light curves which further supports our moving starspot interpretation of the observed periodicities.  

%Our result therefore indicates that assuming 
%This effect shows the complexity of these type of systems, especially when it is expected that systems with short orbital periods are completely synchronised \citep{Fleming19}. 
%\citep{Notsu17} 
%---- no evidence of Balmer emission lines in active G-type stars. 
%\citep{Frasca08} --- emission lines in Ca\,II and H$_{\alpha}$ of a G-type star, but no evidence of emission in other lines.  
%0.82067  

\subsubsection{TYC~3858-1215-1}

TYC\,3858 has been covered in 4 \textit{TESS} sectors, hence, a large amount of data is available for this target.  In contrast to TYC\,110, the \textit{TESS} period of TYC\,3858 is entirely consistent with the spectroscopic period in all four sectors. However, as illustrated in Fig.\,\ref{fig:periodogramTESS}, apart from the orbital period, we also find a strong signal (with several peaks close to each other) at half the orbital period. In what follows we investigate the origin of these signals. 

 We first divided the {\it TESS} light curve in two parts according to their shapes and times when the data was taken, one for sectors $15 + 16$ and one for $22 + 23$. We then fitted a sinusoid with the spectroscopic period to each pair of sectors and removed it from the light curve. We clearly see sinusoidal variations in the residuals but with a different period, i.e. $0.82055(12)$  and $0.82202(88)\,$days for sector $15\,+\,16$ and $22\,+\,23$, respectively, and with a significant phase shift between the two parts of the light curve.  
The residuals are plotted together (middle panel in  Fig.\,\ref{fig:elipse}) phase folded over their period. 

Given that several peaks are seen in the power spectrum close to half the orbital period,
we proceeded by subtracting the sinusoidal signal observed in the residuals and again investigated the 
residuals. The result (Fig.\,\ref{fig:elipse} bottom panel) shows a light curve with a period exactly half the orbital period (0.821100(75)\,days) with reasonable phase alignment. The relative amplitudes for both sectors agrees within the errors.
%We therefore think that this signal might be produced by ellipsoidal variations. 
%The result (Fig.\,\ref{fig:elipse} bottom panel) shows a light curve with a period exactly half the orbital period (0.8211\,days) with reasonable phase alignment with the orbital period. Also the relative amplitudes obtained for both sectors agree within the errors. 

In what follows we interpret the above described findings in the light of the system parameters derived previously. First, the signal at nearly half the orbital period obtained after subtracting the main signal (which corresponds to the orbital period) cannot be caused by ellipsoidal variations. 
The observed change in period between sector $15 + 16$ and $22 + 23$ might appear relatively small but it is statistically significant and, more importantly, the strong phase shift between both light curves cannot be explained by the uncertainty of the measured orbital period. This phase shift excludes ellipsoidal variations as the origin of the observed signal. 
This conclusion is further strengthened by the amplitude of the signal,  $0.90\pm0.01$ and $0.87\pm0.03$ percent for sectors $15 + 16$ and $22 + 23$, respectively, which is far larger than the amplitude predicted 
by Eq.\,\ref{eq:dff}. Even if we let the white dwarf mass vary between $0.2$ and $1.4\mathrm{M}_{\odot}$, the predicted amplitude ranges from  $0.018$ to $0.092$ percent
which is more than an order of magnitude below the measured amplitude. It is impossible to match the measured amplitude without significantly increasing the radius of the main-sequence star (by a factor of 2.1), which would be incompatible with the SED fit and Gaia parallax. 

Given that %the additional signal found after having subtracted the signal at the orbital period and the first sinusoidal at roughly half the orbital period (i.e the signal in the residuals of the residuals) 
the signal in the residuals after subtracting the two main periods is found at half the orbital period, both amplitude and period do not significantly vary between the different sectors, and there is no significant phase shift, one might think that this signal now is finally caused by ellipsoidal variations. 
However, the  amplitude of the observed signal ($0.90\pm0.003$ percent for sectors $15 + 16$ and   
$0.85\pm0.004$ percent for the sectors $22 + 23$) remains an order of magnitude larger than predicted for ellipsoidal variations. Again, this difference cannot be explained by the uncertainties in the stellar or binary parameters. 

We propose a different explanation which, as in the case of TYC\,110, is related to the activity of the secondary star.
%The period difference from one sector to another at nearly half the orbital period, can be explained 
Assuming that two large star spots (or several smaller ones) are located at roughly opposite sides of the secondary star, one would expect brightness variations close to half the orbital period and, as these star spots might slightly move and/or appear/reappear at different latitudes on the K-star surface, one would also expect the obtained periods to somewhat vary with time. 
This scenario could also explain why we see a large number of periods close to half the orbital period in the power spectrum (Fig.\,\ref{fig:periodogramTESS}). In addition, the idea is not in conflict with the measured amplitudes. We therefore conclude that, while it is difficult to provide definitive prove, a reasonable interpretation of the signals at half the orbital is that they are related to the spatial distribution of star spots on the surface of the active secondary star.   

\begin{figure}
	\includegraphics[width=0.78\columnwidth]{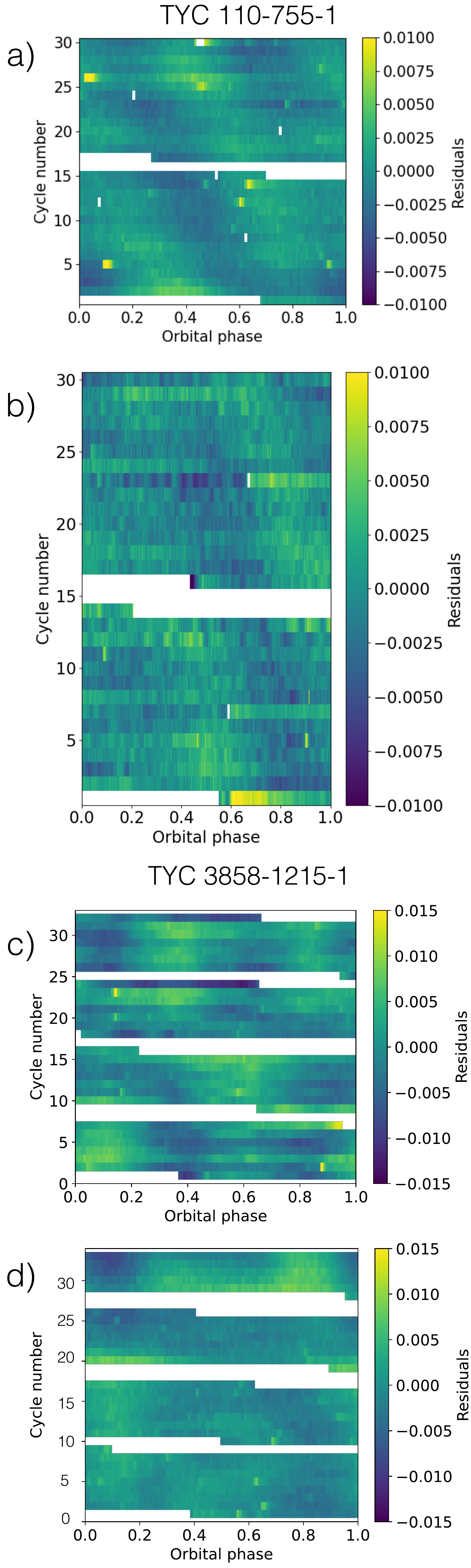}
    \caption{Evolution of the light curve of TYC\,110 (upper panel, {\it a} and {\it b} belongs to sector 5 and 32, respectively) and TYC\,3858 (bottom panel, {\it c} and {\it d} belongs to sector 15+16 and 22+23, respectively) to illustrate the spots migration. The average light curve has been subtracted; the plot shows the residual light curve for each cycle corresponding to the light curve folded on the orbital period.}
    \label{fig:trailerphot}
\end{figure}

This overall interpretation is consistent with the trailed plot (Fig. \ref{fig:trailerphot}) which shows that there are plenty of residual patterns, but these do not move in phase as in the case of TYC\,110. The variations can be associated with new spots appearing at a different place on the surface of the K-type secondary star, rather than the same spots moving. While there is no clear evidence of differential rotation in the light curve of TYC\,3858, the system clearly shows flare activity in agreement with our interpretation of the variations in the light curve shape and the periodic signals found close to half the orbital period.  

%The fact that the optical flares are tiny, is that the extra optical light from the activity is a small fraction of the intrinsic luminosity of a AFGK-type star.  

\subsection{The white dwarfs} 

\begin{figure*}
	\includegraphics[width=1.8\columnwidth]{ 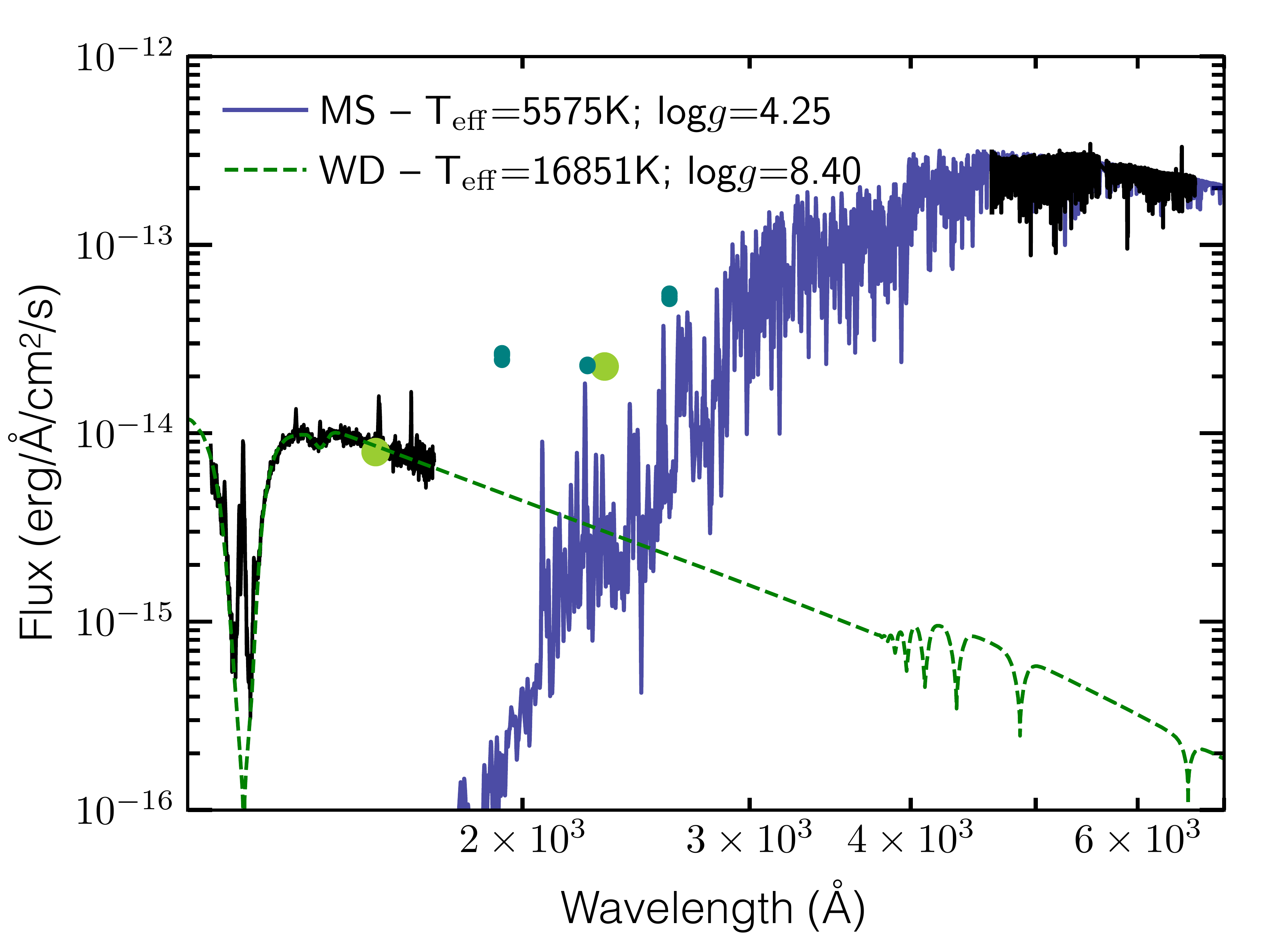}
    \caption{The SED of TYC\,110-755-1 from ultraviolet to optical wavelength ranges. The black lines are the observed {\rm UV}  {\it HST/STIS} spectrum and optical UVES spectrum (no telluric correction was applied). The best-fitting BT-NextGen main-sequence star model is shown in dark blue and was rotationally broadened to match the measured value for the main-sequence star in TYC\,110. The light-green dots represent $FUV$ and $ NUV$ {\em GALEX} observations and in small teal dots UVOT observations from {\it Swift}. We also plot with a green dashed line a pure-hydrogen synthetic model for a $T_{\mathrm{eff}}$=16\,851\,K and $\log{g}$=8.40\,dex white dwarf. 
    We note that the combined flux of our white dwarf model and the model for the secondary star fail to reproduce the NUV measured with  {\em GALEX} and {\it Swift}. We assume that model of the G-type secondary star is underpredicting the NUV emission as chromospheric emission is not taken into account but likely occurring in active G-type stars. 
    }
    \label{fig:SED}
\end{figure*}

We applied the MCMC technique to fit the \textit{HST}/STIS spectrum of TYC\,110 using the {\sc emcee} python package \citep{Foreman13}. 
We started by generating a grid that spans from 9000--30\,000\,K (with steps of 200\,K) and 7.00--8.50 (in steps of 0.1) for the effective temperature and the surface gravity of pure hydrogen atmosphere white dwarfs \citep{koester10-1}. The mixing length parameter for convection was set to 0.8. 
Then we fitted the \textit{HST} spectrum under two assumptions (1) by including
a BT-NextGen model for the secondary star and (2) by ignoring possible contributions from the secondary. We found that the 
results do not differ which confirms that the secondary does not contribute to the far UV emission. Henceforth we consider just a white dwarf model to fit the \textit{HST} spectrum.

For the reddening we also used two approaches. 
First, we assumed the same prior as in the case of the secondary star (restricting the reddening according to the range of values provided by Stilism\footnote{https://stilism.obspm.fr/}
and second we assumed the fixed value taken from the Gaia/2MASS map, i.e.   
E($B-V$)=0.0048\,mag \citep{Lallement19}. The latter value is at the low end of the range of values obtained from fitting the secondary star (which basically does not constrain the reddening, see Fig.\ref{fig:corner}). 
The values derived for the white dwarf parameters using the different approaches concerning the reddening differ by less than eight percent which does not affect any of our conclusions. In Table\,\ref{tab:parameters} we list the values obtained from assuming the lower value of the reddening which we consider more realistic given the distance and location of the target. %

We set a flat prior on the parallax using the EDR3 {\it Gaia} parallax of 7.44\,mas. In contrast, the surface gravity and effective temperature are constrained to be within the limits of the synthetic grid. We set 100 walkers to sample the parameter space with 5000 steps to assure convergence of the chains. The likelihood function to maximize was defined as $-0.5 \chi^2$. The best fit parameters are found by using the median of the marginalized distributions of the parameters and their errors:
 $\Pi =7.454(1)$\,mas, $T_{\mathrm{eff}}=16\,851(35)$\,K, $\log{g}=8.396(8)$\,dex.
The best fit model spectrum is shown in Fig.\,\ref{fig:SED} together with the spectrum and model of the secondary star. 

The mass and the radius of the white dwarf are a function of the surface gravity and effective temperature through the mass-radius relation for white dwarfs. We used the mass-radius relation for white dwarfs with hydrogen-rich atmospheres by interpolating the cooling models from \cite{Fontaine01} with thick hydrogen layers of M$_{\rm H}/$M${_\mathrm{WD}}=10^{-4}$.
We used the last 100 steps of the effective temperature and surface gravity chains to sample the mass-radius relation which does not depend significantly on the assumed thickness of the atmosphere. This produces two distributions (one for the mass and another for the radius). The best values of mass and radius for the white dwarf were extracted with the marginalized distributions and we obtained:  M$_{\mathrm{WD}}=$\,0.784(3)\,M$_\mathrm{\odot}$  and R$_{\mathrm{WD}}=$\,0.0107(4)\,R$_\mathrm{\odot}$.

In the case of TYC\,3858 the presence of a white dwarf has not yet been unambiguously confirmed. However, assuming the {\rm UV} excess coming from a white dwarf companion appears reasonable given that so far all 11 WD+AFGK candidate systems that have been observed with {\it HST}  indeed contained a white dwarf. 
%It is only a possible outcome of the UV emission excess observed with %\te{\it  {\it HST} } observation program, were 100 percent of the observed systems confirm the presence of a WD. 
Knowing the secondary mass and the orbital period, it is possible to estimate the minimum white dwarf mass (0.147 and 0.214\,M$_{\odot}$ for TYC\,110 and TYC\,3858, respectively) if we assume an orbital inclination of 90 degrees. Additionally, it is known that in binaries with short orbital periods the secondary is tidally locked \citep{Fleming19}. A range of possible white dwarf 
masses (0.497--0.597\,$M_{\odot}$ for TYC\,110 and 0.214--0.676\,$M_{\odot}$ for TYC\,3858) can therefore be estimated using the binary mass function based on $v\sin{i}$ and radius measurements of the secondary star. However,
one has to keep in mind that we derive the secondary star parameters by fitting single main sequence star model spectra to observed spectra of  Roche-distorted secondary stars in close binaries, and therefore the white dwarf masses derived from the radius and $v\sin{i}$ have to be considered as rather rough estimates.  
%. 
%This might imply that 
%as our measurements are potentially affected by systematic errors that most likely exceed the purely statistical errors we provide. 

Indeed, comparing the white dwarf masses derived from the  {\it HST}  spectrum of TYC\,110 with our estimate based on $v\sin(i)$, we find a strong disagreement which clearly illustrates that the latter estimate is very uncertain. 
All estimated masses, derived from {\it HST} spectral fitting, the minimum white dwarf mass, and the estimate based on $v\sin{i}$ and synchronised rotation, are shown in Table. \ref{tab:parameters}. 

%With TYC\,110, we have the opportunity to test the accuracy of the binary mass function method, looking at the WD mass obtained with $UV$ spectral fitting. Where the latest is out of the range we provide with the BMFM.

\begin{table}
%	\centering
	\caption{Stellar and binary parameters obtained for the two PCEBs presented in this work. The values provided for the reddening represent the result obtained from the fitting of the secondary star (assuming a prior corresponding to the range of values obtained from \citet{Lallement19}). For the fitting of the HST spectrum of the white dwarf in TYC\,110 we assumed the value provided by the {\it Gaia/2MASS} extinction map which is consistent with the range of values obtained from the fit of the secondary star but at the lower end (which we consider the most realistic assumption). 
}
	\label{tab:parameters}
	\resizebox{\columnwidth}{!}{\begin{tabular}{lll} % four columns, alignment for each
		\hline
           Parameter &  TYC\,110-755-1 & TYC\,3858-1215-1\\
          		\hline
           Orbital Period [days] & 0.85805 $\pm 0.00001$ & 1.6422 $\pm 0.0008$ \\
           Phase zero [BJD] & 2458544.287 $\pm$ 0.003  &  2458404.262 $\pm$ 0.003 \\
           E[$B-V$] & 0.014 $\pm 0.13$ & 0.013$\pm 0.012$  \\
           Distance [pc] &  134.24 $\pm 0.34$ & 68.43 $\pm 0.88$ \\
           Inclination [deg] & 18.67 - 21.49   &  25.33 - 86.64\\
           Sec. Amplitude [km\,s$^{-1}$] & 34.21 $\pm 0.80$  &  44.28 $\pm$ 0.40\\
           Sec. $v\sin{i}$ [km\,s$^{-1}$] & 22.42 $\pm 1.51$  & 27.0 $\pm 18.0$ \\
           $V_\gamma$ [km\,s$^{-1}$]&  -7.6 $\pm 0.6 $ & -17.4  $\pm 0.3 $ \\
           Sec. Luminosity [L$_{\odot}$]& 1.065 $\pm 0.006$ & 0.156 $\pm 0.001$ \\
           Sec. log g [dex] & 4.24 $\pm 0.05$ &  4.55 $\pm 0.04$ \\
           Sec. Z [dex] &  -0.14$\pm 0.14$ & 0.15$\pm 0.37$ \\
           Sec. $T_{\mathrm{eff}}$ [K] & 5562.83 $\pm 13.15$ & 4412.31 $\pm 7.53$ \\
           Sec. Radii [R$_{\odot}$] & 1.114 $\pm 0.006$& 0.679 $\pm 0.004$ \\
           Sec. Mass [M$_{\odot}$] & 0.80 $\pm 0.09$ & 0.61 $\pm 0.07$ \\
           Sec. type & G4V & K3VI\\
           WD minimum Mass [M$_{\odot}$] & -  & 0.214 \\  
           WD Mass(i) [M$_{\odot}$] & -  & 0.214 - 0.676\\ 
           $\ast$WD Mass [M$_{\odot}$] & 0.784 $\pm 0.03$ & - \\
           $\ast$WD $T_{\mathrm{eff}}$ [K] & 16\,851 $\pm 35$ & - \\
           $\ast$WD log g [dex] & 8.39 $\pm 0.00$ & - \\
           $\ast$WD Radii [R$_{\odot}$] & 0.0107 $\pm 0.0004$ & - \\
		\hline
	\end{tabular}}
	
	         We obtained the phase zero, amplitude and $v\sin{i}$ from the radial velocities measurements.\\
	         $V_{\gamma}$ is the radial velocity of the center of mass of the system.\\ 
	         The white dwarf Mass(i) corresponds to $v\sin{i}$ method.\\
			 $\ast$ {\it HST}  spectra fitting. 
\end{table}

\section{Scrutinizing the unusual emission lines in TYC~110-755-1}\label{sec:lines}
%we tested possibility  and detected no significant variability

\begin{figure}
    \centering
    \includegraphics[width=\columnwidth]{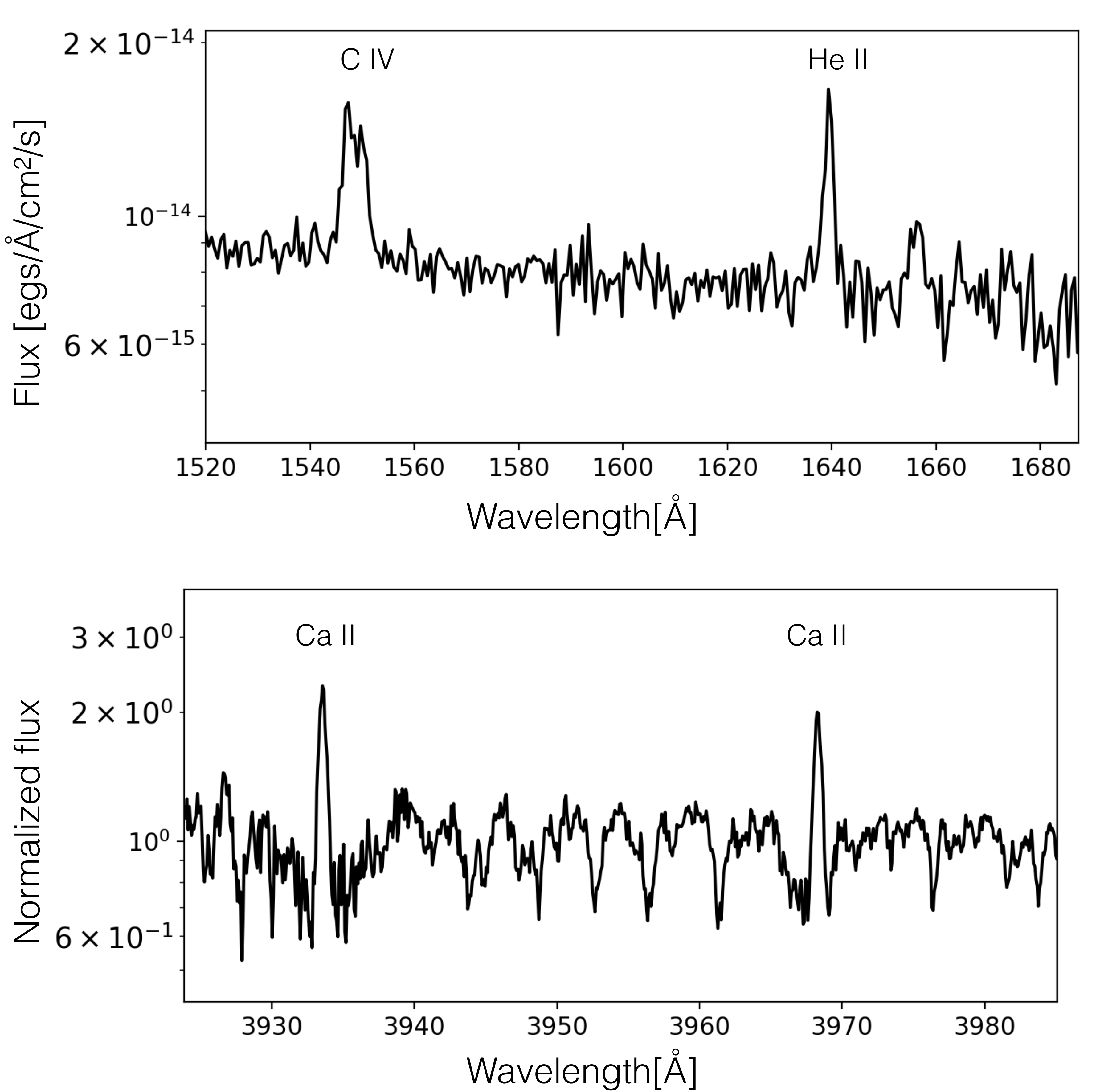}
    \caption{Two small portions of the full spectra of TYC\,110-755-1. The strong He\,II line at 1640\,\AA\, and C\,IV at 1550\,\AA\, observed in the {\it HST} spectra are shown in the upper panel, while the chromospherical emission lines of  Ca\,II at 3933.66 and 3968.47\,\AA\, in the optical range are shown in the bottom panel.}
    \label{fig:spec}
\end{figure}

As we have learned in the previous  sections, TYC\,110 shows multiple signs of chromospheric/magnetic activity typical for G-stars such as spots, flares, and now chromospheric emission lines such as  H$_{\alpha}$, and Ca\,II (Fig.\,\ref{fig:spec}). However, in addition to these rather typical features strong He\,II line at 1640\,\AA\, and C\,IV at 1550\,\AA\, (see Fig.\,\ref{fig:spec}) are found in the {\it HST}-{\rm UV} spectrum.  
%although there is no evidence for Roche Lobe overflow in TYC\,110, and no accretion at the current stage of evolution is foreseen 
%The spectrum shows. 
While the most natural explanation is that these emission lines are related to activity as well, for completeness we tested 
%are quite unusual, even for cataclysmic variables \citep{Long05}. 
the alternative possibility that wind accretion onto a magnetic white dwarf is the origin of the {\rm UV}-emission lines as accretion onto the magnetic poles could produce such emission lines.% if TYC\,110 contains a strongly magnetic WD.

To that end, we observed TYC\,110 using {\it Swift} orbital telescope as a target of opportunity (ToO target ID:14068) from February 22 to March 7, 2021. One orbit exposures were obtained per day to cover at least two orbital periods, with a total on-source exposure time of  23\,165\,sec. The average count rate is 0.043 counts per second. Sixteen {\it Swift} orbits covered all orbital phases of the object evenly. We used the X-ray Telescope \citep[XRT,][]{Burrows05} and the Ultraviolet/Optical Telescope \citep[UVOT,][]{Roming05} to analyse data for the presence of possible pulses produced by a spinning magnetic white dwarf.
Standard data processing was made later at the {\it Swift} Science Data Centre. The data were collected with the XRT operating in AUTO mode. Simultaneous {\rm UV} images in the $UVW1$, $UVW2$ and $UVM2$ at 2600, 1928 and 2246\,\AA, respectively, were obtained with the UVOT. We used an on-line tool\footnote{https://www.swift.ac.uk/} to extract the XRT light curve and standard aperture photometry using Web-HERA provided by The High Energy Astrophysics Science Archive Research Center\footnote{https://heasarc.gsfc.nasa.gov/docs/tools.html} \cite[HEASARC,][]{Blackburn99}, corresponding procedures to measure {\rm UV} fluxes. In both cases HEASOFT v6.28 software and calibration were used.  
 
No significant phase-dependent variability was detected, neither at X-ray nor at {\rm UV} wavelengths. Hence, we omit unnecessary figures to present these data. However, {\rm UV} measurements (two for each $UVW1$, $UVW2$ and $UVM2$ bands) are presented in Fig.\,\ref{fig:SED} (small dark green dots). We note that the flux in the NUV predicted by the models obtained from our spectral fitting is not sufficient to reproduce the observations. We assume that this is most likely a consequence of chromospheric emission of the secondary star 
in the NUV which is not considered in the models.

Despite being difficult to prove, this hypothesis appears reasonable as large differences between predicted and observed NUV fluxes for active stars have been reported previously. For example, the chromospheres of active M dwarfs can fully dominate their NUV fluxes, i.e. the photospheric flux predicted by model spectra is underestimating the NUV flux of these stars by orders of magnitude \citep[][their figures 6 and 7]{stelzeretal13-1}. In addition, for evolved stars a relation between excess NUV flux and rotation has been found \citep{dixonetal20-1} and for fast rotators the NUV flux from the chromosphere exceeds that of the photosphere by up to four magnitudes \citep[see also][their figure 19]{jayasingheetal21-1}. 

While these results are not directly applicable to the G type secondary star in TYC\,110, given the fast rotation of the star and the fact that is showing strong signs of activity in its light curve, it is plausible that steady chromospheric emission which is not incorporated in the model spectra is responsible for the large discrepancy between our fits and the observations in Fig.\,\ref{fig:SED}.
%A much less likely alternative would be that a UV bright background AGN is responsible for the missing NUV flux, a situation which might have occurred in a different target studied by us 
%(Lagos et al. 2022, MNRAS, submitted) where we discovered a background galaxy using high contrast imaging. 
%Distinguishing between these two possibilities is beyond the scope of the paper especially because the explanation through cromospheric emission is plausible. }}

%While $FUV$ is completely dominated by the white dwarf, we noticed there is a $NUV$ excess, produced by the enhanced chromospheric contribution of very active main--sequence stars as it was seen in \citet[][Fig 12]{Loyd16}. The chromospheric contribution vary on long time scales, resulting unlikely to detect the variation with our observations.}

As the {\it Swift} observations did not confirm the magnetic white dwarf hypothesis, we conclude that the system is a PCEB with a particularly active secondary star and that the emission lines are a result of this activity. 
%TYC\,110 can be compared with the RS Canum Venaticorum variable \citep[RS~CVn,][]{Rodono87},
In fact, in certain aspects the system resembles V471\,Tau \citep{Sion12} which also contains a rapidly rotating main sequence K-star with spots and flare activity. V471\,Tau also shows emission lines in the {\rm UV}, including C\,IV and He\,II, and is an X-ray source.

\section{Past and future evolution}\label{sec:past}

Previous attempts to understand the formation and evolution of close white dwarf binaries, particularly with massive companions (AFGK),  indicate that, to unbind the common
envelope from the binary, the energy needed might be higher than in those systems with M-dwarf companions %indicate that the energy needed to unbind the common envelope from the binary might be higher than the energy needed for those systems with M-dwarf companions 
\citep[e.g.,][]{zorotovicetal14-1,zorotovicetal14-2}. However, in recent papers we presented four systems with G-type secondaries that do not support that hypothesis (\citetalias{parsons16}, \citetalias{Hernandez21}). 
%The difference then, might rely on the orbital period of the system. 
%\subsection{Past and future evolution}

With the aim of further constraining theories of close compact binary star evolution, 
we have reconstructed the past evolution of the two systems following \citetalias{Hernandez21}, i.e. using the algorithm developed by \citet{zorotovicetal10-1}, leaving the common envelope efficiency ($\alpha_{\mathrm{CE}}$) as a free parameter and assuming that no other sources of energies (like the hydrogen recombination energy) contributed to the ejection process.
Table\,\ref{tab:rec} lists the reconstructed parameters of the possible progenitors for both systems. The first row for each system corresponds to the reconstructed parameters assuming the same efficiency derived for PCEBs consisting of a white dwarf and an M-dwarf companion from SDSS (i.e., $\alpha_{\mathrm{CE}} = 0.2-0.3$, \citealt{zorotovicetal10-1}), while the second row is for the whole range of possible efficiencies. For TYC\,110 we provide the results %for two white dwarf masses, the one we derived from the mass function and 
 for the more reliable mass of the white dwarf based on {\it HST} spectral fitting.
As for the three systems studied in \citetalias{Hernandez21}, TYC\,110 and TYC\,3858 do not require extra energy sources, apart from the available orbital energy, to have survived a common-envelope phase, regardless of the set of derived parameters assumed for the reconstruction of TYC\,110. Given their short current orbital periods, %both systems
TYC\,3858 can be reconstructed with virtually any value of $\alpha_{\mathrm{CE}} \leq 1$ %with the exception that the parameters derived 
while for TYC\,110 and the {\it HST} spectral fitting imply $\alpha_{\mathrm{CE}} \geq 0.19$.  % allowing for a large range of initial primary masses, initial orbital periods and ages at the onset of the common envelope phase. 
For TYC\,110 we found possible progenitors for the whole range of derived white dwarf masses. %, for both fitting methods. For the parameters derived from the mass function, the reconstruction suggests that the primary star filled its Roche lobe either on the early or the thermally pulsating asymptotic giant branch, while
The parameters derived from the {\it HST} spectral fitting imply a  massive progenitor that filled its Roche lobe during the thermally pulsing asymptotic giant branch phase when the system was very young. 
For TYC\,3858, on the other hand, the common envelope reconstruction imposes a lower limit of $0.27\,M_\odot$ on the white dwarf mass, and the progenitor of the white dwarf may have filled its Roche lobe either on the first giant branch or on the asymptotic giant branch. 

\begin{table}
	%\centering
	\caption{Reconstructed binary parameters assuming common envelope evolution. $M_{\mathrm{1,i}}$ and $P_{\mathrm{orb,i}}$ are the initial primary mass and initial orbital period, respectively. For each system, the first row lists the parameters constraining $\alpha_{\mathrm{CE}}$ in the range of 0.2-0.3, while the second row is for  any $\alpha_{\mathrm{CE}}\leq1$.}
	\label{tab:rec}
	\resizebox{\columnwidth}{!}{\begin{tabular}{llcccc} % four columns, alignment for each
		\hline
	Object & $\alpha_{\mathrm{CE}}$ & WD Mass [M$_\odot$] &  $M_{\mathrm{1,i}}$ [M$_\odot$] & $P_{\mathrm{orb,i}}$ [days]& CE Age[Gyr] \\
	%  & & & initial & initial \\ 
	\hline
	%TYC 110  & 0.2--0.3  & 0.52--0.65 & 1.65--2.71 & 313--1108 & 0.64--2.31\\
	 %        & 0.04--1.0 & 0.52--0.65 & 1.24--2.87 &  59--1662 & 0.54--5.81\\
   % &  & & & & \\
	TYC 110 ( {\it HST} )  & 0.2--0.3  & 0.75--0.82 & 3.24--3.71 & 929--1947 &                          0.26--0.38\\
	                & 0.19--1.0 & 0.75--0.82 & 3.15--3.75 &  147--2015 &     
	                  0.26--0.42\\
	&  & & & & \\
    TYC 3858 & 0.2--0.3  & 0.34--0.62 & 0.98--2.11 & 103--1477 & 1.29--13.29 \\
             & 0.03--1.0 & 0.27--0.68 & 0.98--2.98 &  22--1591 & 0.49--13.29\\
       \hline
	\end{tabular}}
\end{table}

%\subsection{The future} 

WD+AFGK binaries are crucial for our understanding of close white dwarf binary formation not only because these systems constrain common envelope evolution theories 
but also because we can estimate the future evolution of these systems using detailed simulations.  %Their total mass can exceed the Chandrasekhar limit and therefore, in contrast to most WD plus M-dwarf binaries, some of these systems might potentially be SN\,Ia progenitors.  
%as the total mass of the binary usually exceeds the 
%A quality of PCEBs with solar type or more massive secondaries in contrast with WD+M-dwarf PCEBs, is the 
%A difference between solar type and late-type secondary stars in PCEBs lies in the 
%diversity of the pathways which they can evolve. 
While white dwarfs with late-type secondary stars will all become cataclysmic variables when the secondary fills its Roche lobe, the future of PCEBs with solar type secondary stars depends on several factors such as the orbital period at the end of common envelope evolution, the angular momentum loss mechanism that drives the system to shorter orbital periods, the mass ratio of the binary, and the evolutionary status of the secondary star.

The three systems shown in \citetalias{Hernandez21} provide evidence for the diversity of evolutionary futures of close WD+AFGK binaries and we here further confirm this finding. TYC\,110 contains a G-type secondary star like in the cases of TYC\,4700-815-1, TYC\,1380-957-1 and TYC\,4962-1205-1, while TYC\,3858 contains a K-type secondary star, and all these binaries have orbital periods between 0.85 and 2.5 days. 
Despite their similarities, the future of these systems are quite different. 
With the aim to predict how these systems will evolve and what kind of interacting binary stars they will form in the future, we performed dedicated simulations using \textsc{MESA} \citep{paxton11}, with the same set-up and the same criteria for stable mass transfer as in \citet{parsons15} and \citetalias{Hernandez21}. For each system we adopt the parameters listed in Table\,\ref{tab:parameters}.

%\subsubsection{TYC\,110-755-1}

\begin{figure}
	\includegraphics[width=\columnwidth]{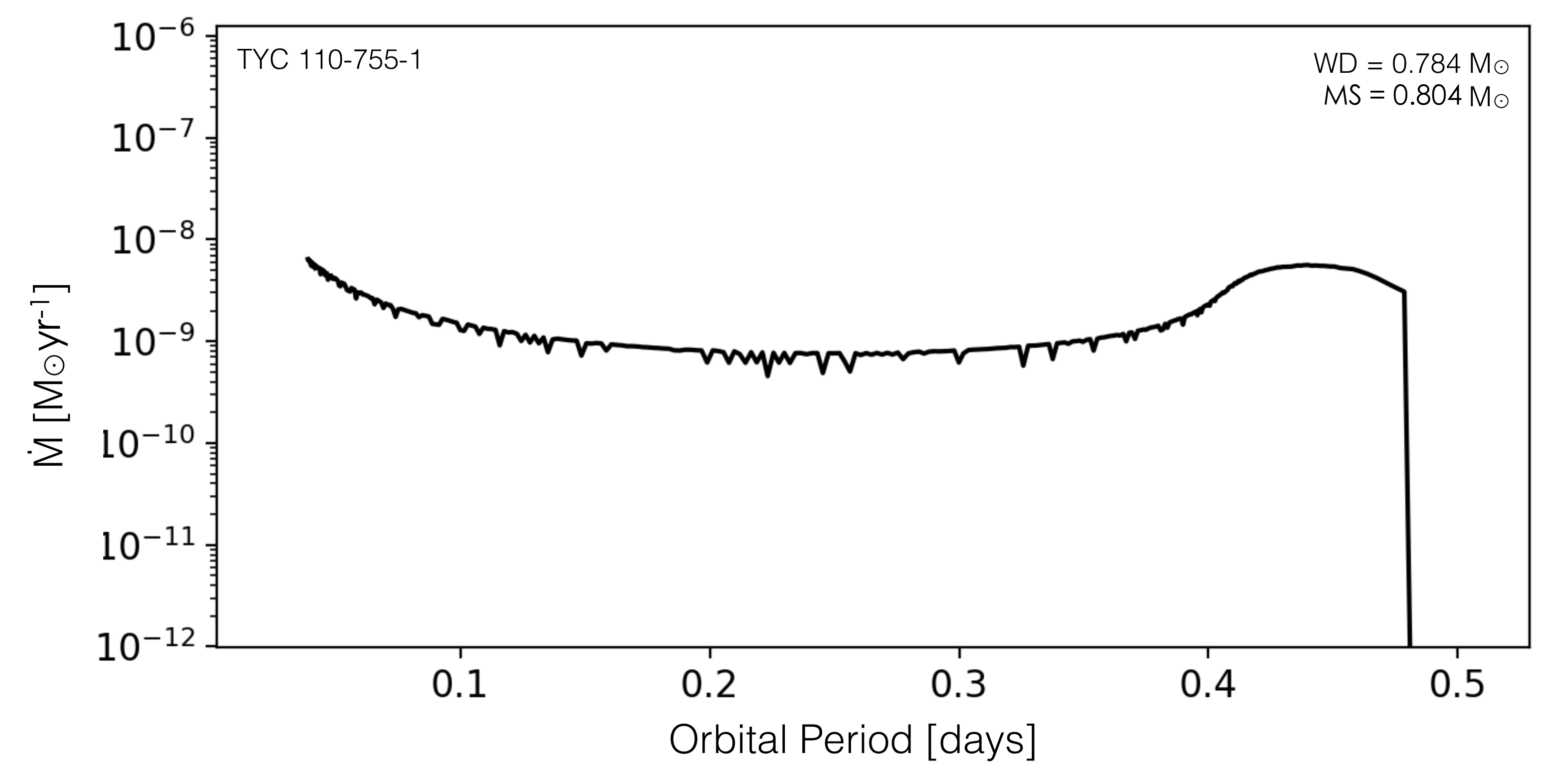}
    \caption{The future evolution of TYC\,110-755-1 for a white dwarf with the {\it HST} mass estimation of 0.78\,M$_{\odot}$ and a main sequence star of 0.804\,M$_{\odot}$. This system will become a cataclysmic variable with an evolved donor star. Mass transfer mainly driven by magnetic breaking will continue during the orbital period range of 2-3\,hr. }
    \label{fig:future}
\end{figure}

Simulations for TYC\,110 were performed assuming a secondary mass of M$_{\mathrm{MS}}=0.804$\,M$_{\odot}$, a white dwarf mass of M$_{\mathrm{WD}}=0.78$\,M$_{\odot}$, the current orbital period of the system $\mathrm{P_{orb}}=0.858$\,days  and the current radius of the secondary star of R$_{\mathrm{MS}}=1.114$\,R$_{\odot}$. The radius of TYC 110 is larger than that of a star of the same mass on the zero age main sequence (ZAMS), which indicates that the secondary is slightly evolved. The starting model for our MESA simulations was obtained by evolving the secondary until it reaches the required radius of the evolved secondary star. Due to the large white dwarf mass, mass transfer remains dynamically and thermally stable and the system evolves into a CV. In 15\,Myr from now, when the orbital period has been reduced to 11\,hrs, mass transfer starts but never reaches values large enough to ignite stable nuclear burning on the surface of the white dwarf. In this case nova eruptions should occur and the mass of the white dwarf remains nearly constant (non-conservative mass transfer). Thus, TYC\,110 will evolve into a CV with a significantly evolved secondary.
Driven out of thermal equilibrium due to mass transfer, the evolved secondary does not become fully convective and angular momentum loss through magnetic braking will be efficient leading to large mass transfer rates until it reaches the orbital period minimum of $\sim\,0.932$\,hrs in 57\,Myr. The system will not evolve as a detached system through the orbital period gap (see Fig. \ref{fig:future}).
This evolutionary path for cataclysmic variables with evolved secondary stars has been predicted by \citet{Kolb+Baraffe00} and several such CVs have been found \citep[e.g.][]{thorstensenetal02-1,rebassa-mansergasetal14-1}. 
%While TYC\,4962 is the first known progenitor of a CV with an evolved donor, TYC\,110 is the second known system.

The future of TYC\,3858 is straightforward to predict. Given the mass ratio of  0.79 < $q$ < 3.17 where $q=M_{MS}/M_{WD}$, the secondary mass  of $0.61\,M_{\odot}$ and radius of $0.68\,R_{\odot}$ the system will run into dynamically unstable mass transfer when the secondary fills its Roche-lobe in 2\,Gyr from now, when the orbital period will be 6 hrs. This is predicted by our MESA simulations in agreement with \citet[][]{geetal15-1}. The orbital energy of the binary will not be sufficient to expel the envelope and therefore the two stars will merge and evolve into a giant star with the core of this giant being the white dwarf we observe today. This giant will then evolve into a single white dwarf as the final fate of TYC\,3858.

In Table\,\ref{tab:periodlist} we list the orbital periods of all PCEBs with secondary stars earlier than M. Thanks to our survey a population of systems with short orbital periods (below 1-2 days) could be established. At the same time, a population of systems with significantly longer orbital periods (weeks to years) exists. If these indications for a bimodal distribution can be confirmed due to further observations, the population of WD+AFGK binaries might consist of two sub-samples that perhaps formed through different evolutionary channels.

\begin{table}
\setlength{\tabcolsep}{4pt}
    \caption{list of known WD+AFGK PCEBs sorted by orbital period. We include the two systems described in this work.}
    \centering
    \begin{tabular}{llllc}
    \hline
        Object & P$_\mathrm{orb}$ [days]  & M$_\mathrm{sec}$ [M$_\odot$] & M$_\mathrm{WD}$ [M$_\odot$] & Reference\\
        \hline
        GPX-TF16E-48 & 0.2975 & 0.64 & 0.72 &  7\\
        TYC\,6760-497-1 & 0.4986 & 1.23 & 0.52--0.67 & 5 \\
        V471\,Tau &  0.5208 & 0.93 & 0.84 & 6 \\
        TYC\,110-755-1 & 0.858 & 0.84 & 0.784 & 9\\
       % *V1082\,Sgr & 0.867 & 0.73& 0.64 & 6\\
        TYC\,4962-1205-1 & 1.2798 &  0.97 & 0.59--0.77 & 8 \\
        TYC\,1380-957-1 & 1.6127 & 1.18 & 0.64--0.85 & 8 \\
        TYC\,3858-1215-1 & 1.6422 & 0.61 & 0.22-0.68 & 9\\
        TYC\,4700-815-1 & 2.4667 & 1.45 & 0.38--0.44 & 8\\
        IK\,Peg & 21.722 & 1.70 & 1.19 & 1\\
        KOI-3278 & 88.1805 &  1.05 & 0.634	 & 2\\
        SLB3 & 418.72 & 1.56 & 0.62 & 3\\
        KIC\,8145411 & 455.87 & 1.12 & 0.197 & 4\\
        SLB1 & 683.27 & 1.106 & 0.53  & 3\\
        SLB2 & 727.98 & 1.096 & 0.62 & 3\\
        \hline
    \end{tabular}
            1-\citet{Wonnacott93}, 2-\citet{Kruse+agol15}, 3-\citet{Kawahara18}, 4-\citet{Masuda19}, 5-\citet{parsons15},  6-\citet{OBrien01},   7-\citet{Krushinsky20}, 8-\citetalias{Hernandez21},
            9-This work.%6-\citet{Tovmassian18},

    \label{tab:periodlist}
\end{table}

\section{Conclusions}
%Two post common envelope binary systems with solar type secondary stars and short orbital periods were identified and introduced in this work; 
We presented the discovery of two post common envelope binaries (PCEBs) with intermediate mass (G and K spectral type) secondary stars. We used high resolution optical spectroscopy to determine stellar parameters of the secondary star of both systems, and STIS {\it HST} spectroscopy to determine the white dwarf parameters of TYC\,110. 

%We also obtained \textit{TESS} light curves for both systems and find .... 
%TYC\,3858, where the presence of the WD is only indicated by UV excess detected with {\it GALEX} and TYC\,110, which WD was confirmed with {\it  {\it HST} } observations. 
%We here highlight two implications of our results. 
Based on the orbital and stellar parameters obtained from these observations, we reconstruct the past and predict the future evolution of both systems. We find that both systems join the class of close 
WD+AFGK stars that do not require any modifications of the common envelope prescription when compared to PCEBs with M-dwarf companions. Our systematic survey of WD+AFGK 
binary stars has now discovered six such systems (this paper, \citetalias{Hernandez21} and \citet{parsons15}) which clearly indicates that a population of PCEBs with secondary stars earlier than M exists that forms through 
common envelope evolution in the same way as PCEBs with M-dwarf companions. 
Given that reconstructing the evolution of two other close WD+AFGK systems, IK\,Peg and KOI-3278 \citep{zorotovicetal14-1}, does require additional energy sources during common envelope evolution, 
one might speculate that two different populations of WD+AFGK binaries emerge from the first mass transfer phase which, if confirmed, might have far reaching consequences 
for our understanding of close white dwarf binary formation. 

If this finding can be further confirmed, an immediate question that arises is whether the longer orbital period systems are absent among close WD+M binaries. As there is no obvious reason why the common envelope efficiency should be higher for some WD+AFGK binaries and why additional sources of energy should be available for some of them, we speculate that the longer orbital period systems among WD+AFGK stars are perhaps the result of dynamically stable mass transfer. This type of mass transfer requires mass ratios close to one for the main sequence progenitor systems. As these mass ratios are impossible for the progenitors of WD+M binaries, this scenario naturally explains why longer orbital period systems are not found among them. While two populations of post mass transfer WD+AFGK binaries, one with short orbital periods from CE evolution and one with longer orbital periods emerging from stable mass transfer is therefore a tempting interpretation of the results obtained so far, more observations are needed to establish the existence of two different populations.

%If the existence of two different populations of WD+AFGK can 
%While there is no obvious reason to assume 

While the main focus of our survey is to progress with our understanding of white dwarf binary formation and evolution, in this work we learned that 
PCEBs with earlier secondary stars are excellent laboratories for studying magnetic activity. 
Using detailed {\it TESS} light curves of both systems studied here, we find that both systems have rapidly rotating secondary stars and show plenty of flares and starspots. Nevertheless, the difference between both systems can not be more sharp. %In one side,
While TYC\,3858 shows short-lived spots that appear to remain mostly fixed, 
%This would suggest that the $v\sin{i}$ method to measure the WD mass should be ideal for TYC\,3858. Conversely, 
TYC\,110 shows long-lived spots that migrate across a differentially rotating surface.
We also found that {\it TESS} light curves can be used to identify close binaries in our targets sample, which will reduce 
the amount of required spectroscopic follow-up observations.

The future of the two systems we presented in this work will be very different. TYC\,3858 will run into dynamically unstable mass transfer and evolve into a single white dwarf while TYC\,110 will evolve 
into a cataclysmic variable star with an evolved donor star. Combining this finding with the results obtained by \citet{parsons15} and \citetalias{Hernandez21} 
who found systems that evolve into the same channels or into thermal time scale mass transfer (becoming super soft sources for some time) 
we conclude that small changes in the system parameters can drastically change the fate of these systems. This is because, for these earlier spectral type secondary stars, the critical mass ratios 
for stable as well as dynamically and thermally unstable mass transfer are relatively close to each other.    
%Despite all the similarities of this systems, we have learned that PCEB with solar type secondary stars are sensitive to small changes that can produce immensely different pathways in which they can evolve. 
This emphasizes the importance of conducting a WD+AFGK binary survey, since a large sample of observationally well characterized systems can provide 
crucial constraints for white dwarf binary evolution theory with implications for the SN\,Ia progenitor problem.

%it might be that the population of close WD+AFGK binaries consists of two different 
%emerging from the first phase of mass transfer. 

%We found a strong similarity between TYC\,110 and V471\,Tau which is a unique and exciting object; identifying a similar star is important and warrants further study. 

%otra cosa, si hay acrecion (y no es corto flare que causa espectro UV) entonces el continuo en optico sera afectado
%y tu temperatura puede ser equivocada. como pasa en V471\,Tau.

\section*{Acknowledgements}
MSH acknowledges the Fellowship for National PhD students from ANID, grant number 21170070. 
MRS acknowledges FONDECYT for the financial support (grant 1181404).
SGP acknowledges the support of the STFC Ernest Rutherford Fellowship. 
BTG was supported by the UK STFC grant ST/T000406/1.
OT was supported by a Leverhulme Trust Research Project Grant and FONDECYT grant 3210382. 
GT acknowledges support from PAPIIT project IN110619.
MZ acknowledges financial support from FONDECYT (Programa de Iniciaci{\'o}n, grant 11170559). 
FL was supported by the ESO studentship and the National Agency for Research and Development (ANID) Doctorado nacional, grant number 21211306.
RR has received funding from the postdoctoral fellowship program Beatriu de Pin\'os, funded by the Secretary of Universities and Research (Government of Catalonia) and by the Horizon 2020 program of research and innovation of the European Union under the Maria Sk\l{}odowska-Curie grant agreement No 801370. 
ARM acknowledges support from the MINECO grant AYA\-2017-86274-P and by the AGAUR (SGR-661/2017). Grant RYC-2016-20254 funded by MCIN/AEI/10.13039/501100011033 and by ESF Investing in your future.
JJR acknowledges the support of the National Natural Science Foundation of China 11903048 and 11833006.  We acknowledge the use of public data from the {\it Swift} data archive. 

\section*{Data Availability}
The data underlying this article will be shared on reasonable request to the corresponding author.

%%%%%%%%%%%%%%%%%%%%%%%%%%%%%%%%%%%%%%%%%%%%%%%%%%



%\bibliographystyle{mnras}


\typeout{}
\bibliographystyle{mnras}
\bibliography{references.bib}

%%%%%%%%%%%%%%%%%%%%%%%%%%%%%%%%%%%%%%%%%%%%%%%%%%

%%%%%%%%%%%%%%%%% APPENDICES %%%%%%%%%%%%%%%%%%%%%

\appendix

\section{Radial velocities}

\begin{table}
\setlength{\tabcolsep}{4pt}
	\centering
	\caption{TYC 110-755-1 Radial velocity measurements.}
	\label{tab:RV_tyc110}
	\begin{tabular}{llllll} % four columns, alignment for each
		\hline
		Instrument & S/N & Exptime &BJD & RV & error\\
		            & &$s$ &    & $Km~s^{-1}$ & $Km~s^{-1}$\\
		\hline
        ESPRESSO & 30 & 1200  &2458450.8084	& 5.53	& 1.33\\
        ESPRESSO & 28 & 1200  &2458451.7378	& 21.91 & 	0.72\\
        ESPRESSO & 33& 1200 &2458804.7751	& -23.08& 	1.13\\
        ESPRESSO & 34& 1200 &2458176.0456	& -39.90   & 	2.10\\
        ESPRESSO & 29& 1200 &2458857.7013	& 24.18& 	1.51\\
        ESPRESSO & 31& 1200 &2458859.7440	& -32.70& 	0.76\\
        ESPRESSO & 33& 1200 &2458860.7120	& -39.23& 	1.17\\
        ESPRESSO & 30& 1200 &2458861.7381	& -12.33& 	0.91\\
        ESPRESSO & 28& 1200 &2458915.6990 & 	-32.55	& 1.09\\
        ESPRESSO & 29& 1200 &2458916.6745	& -7.18& 	1.03\\
        ESPRESSO & 33& 1200 &2458917.6707	& 20.54& 	0.55\\
		HRS   & 45 &2400& 2458170.0718 &	-37.20 &	1.40 \\
        HRS   & 35 &1000&2458172.032480 &	-5.70  & 4.50 \\
        UVES  & 69 & 120& 2458355.850589 &	15.06	& 0.61 \\
        UVES  & 64 & 120& 2458356.842519 &	-17.95	& 0.64 \\
        UVES  & 69 & 120&2458357.824764&	-41.62	& 0.64 \\
		\hline
	\end{tabular}
\end{table}

\begin{table}
\setlength{\tabcolsep}{4pt}
	\centering
	\caption{TYC 3858-1215-1 Radial velocity measurements.}
	\label{tab:RV_tyc3858}
	\begin{tabular}{llllll} % four columns, alignment for each
		\hline
		Instruments & S/N &Exptime &HJD & RV & error\\
		            &   & $s$&  & $Km~s^{-1}$ & $Km~s^{-1}$\\
		\hline
        ESPRESSO & 26 & 1200  & 2457945.7164 & 	-61.78 & 	0.86\\
        ESPRESSO & 29 & 1200  & 2457946.7218 & 	10.56 & 	0.81\\
        ESPRESSO & 25 & 1200  & 2457947.6950 & 	-22.02 & 	0.80\\
        ESPRESSO & 20 & 1200  & 2457951.6644 & 	8.09 & 	    0.75\\
        ESPRESSO & 30 & 1200  &  2457948.6984 & 	-40.83 & 	0.86\\
        ESPRESSO & 31 & 1200  & 2457953.6797 & 	-53.32 & 	0.90\\
        ESPRESSO & 28 & 1200  & 2458277.7669 & 	-25.55 & 	0.86\\
        ESPRESSO & 25 & 1200  & 2458277.7811 & 	-24.00 & 	0.82\\
        ESPRESSO & 29 & 1200  & 2458278.7998 & 	-40.90 & 	0.74\\
        ESPRESSO & 31 & 1200  & 2458278.8141 & 	-43.20 & 	0.64\\
        ESPRESSO & 30 & 1200  & 2458279.7631 & 	26.40 & 	0.80\\
        ESPRESSO & 27 & 1200  & 2458856.9729 & 	-56.22 &    0.89\\
        ESPRESSO & 24 & 1200  & 2458856.9870 & 	-56.93 & 	0.87\\
        ESPRESSO & 28 & 1200  & 2458857.0012 & 	-57.96 & 	0.87\\
        ESPRESSO & 24 & 1200  & 2458857.0174 & 	-57.61 & 	1.18\\
        ESPRESSO & 29 & 1200  &  2458857.0316 &  -57.54 & 	1.12\\
        ESPRESSO & 31 & 1200  & 2458857.0457 & 	-59.06 & 	1.13\\
        ESPRESSO & 32 & 1200  & 2458857.9501 & 	26.17 & 	0.77\\
        ESPRESSO & 30 & 1200  & 2458857.9642 & 	26.90 & 	0.64\\
        ESPRESSO & 29 & 1200  & 2458857.9784 & 	26.37 & 	0.59\\
        ESPRESSO & 26 & 1200  & 2458858.0227 & 	20.73 & 	0.69\\
        ESPRESSO & 28 & 1200  & 2458858.0368 & 	20.32 & 	0.70\\
        ESPRESSO & 34 & 1200  & 2458858.0510 & 	19.56 &	    0.75\\
        ESPRESSO & 36 & 1200  & 2458859.9596 & 	-14.97 &	0.95\\
        ESPRESSO & 31 & 1200  & 2458859.9737 & 	-16.42 & 	0.93\\
        ESPRESSO & 33 & 1200  & 2458859.9879 & 	-18.09 & 	0.99\\
        ESPRESSO & 29 & 1200  & 2458860.0571 & 	-28.72 & 	0.75\\
        ESPRESSO & 28 & 1200  & 2458860.0713 & 	-31.23 & 	0.83\\
        ESPRESSO & 29 & 1200  & 2458860.9626 & 	10.95 & 	0.87\\
        ESPRESSO & 31 & 1200  & 2458860.9770 & 	13.40 & 	0.93\\
        ESPRESSO & 30 & 1200  & 2458860.9911 & 	15.23 & 	0.92\\
        ESPRESSO & 33 & 1200  & 2458861.0794 & 	17.98 & 	0.80\\
        ESPRESSO & 27 & 1200  & 2458861.9640 & 	-61.94 & 	0.71\\
        ESPRESSO & 29 & 1200  & 2458861.9790 & 	-61.64 & 	0.75\\
        ESPRESSO & 28 & 1200  & 2458861.9932 & 	-60.82 & 	0.73\\
        ESPRESSO & 30 & 1200  & 2458862.0432 & 	-64.90 & 	0.87\\
        ESPRESSO & 31 & 1200  & 2458862.0574 & 	-64.09 & 	0.90\\
        ESPRESSO & 26 & 1200  & 2458862.0717 & 	-64.16 & 	0.91\\
        ESPRESSO & 28 & 1200  & 2458279.7775 & 	26.79 & 	0.84 \\
		\hline
	\end{tabular}
\end{table}

\bsp	% typesetting comment
\label{lastpage}
\end{document}